\documentclass[12pt,preprint]{aastex}

\shorttitle{Wave propagation in magnetic structures}
\shortauthors{Centeno et al.}

\begin{document}

\title{Wave propagation and shock formation in different magnetic structures}

\author{R. Centeno\altaffilmark{1,2}, M. Collados\altaffilmark{2}, J. Trujillo Bueno\altaffilmark{2,3}}
\altaffiltext{1}{High Altitude Observatory (NCAR) The National Center for Atmospheric Research is sponsored by the National Science Foundation, Boulder CO 80301, USA}
\altaffiltext{2}{Instituto de Astrof\'\i sica de Canarias, 38205, La Laguna, Tenerife (Spain)}
\altaffiltext{3}{Consejo Superior de Investigaciones Cient\'\i ficas (Spain)}
\email{rce@ucar.edu, mcv@iac.es, jtb@iac.es}

\begin{abstract}

Velocity oscillations "measured" simultaneously at the photosphere and
the chromosphere -from time series of spectropolarimetric data in the 
10830 \AA\ region- of different solar magnetic features allow us to study 
the properties of wave propagation as a function of the magnetic flux of 
the structure (i.e. two different-sized sunspots, a tiny pore and a facular region). While photospheric oscillations have similar characteristics 
everywhere, oscillations measured at chromospheric heights show different 
amplitudes, frequencies and stages of shock development depending on the 
observed magnetic feature. The analysis of the power and the phase spectra, 
together with simple theoretical modeling, lead to a series of results 
concerning wave propagation within the range of heights of this study. 
We find that, while the atmospheric cut-off frequency and the propagation 
properties of the different oscillating modes depend on the magnetic feature, 
in all the cases the power that reaches the high chromosphere above the 
atmospheric cut-off comes directly from the photosphere by means of linear 
vertical wave propagation rather than from non-linear interaction of modes.

\end{abstract}


\section{Introduction}

Encrypted in the oscillatory behavior of the solar atmosphere lies crucial information
for understanding its dynamical and physical properties. The stratification caused by gravity, together with the presence of
magnetic fields, leads to a variety of magneto-gravity-acoustic modes \citep{khomenko2006, abdelatif, thomas}.
By analyzing the local properties of oscillations and wave propagation we can 
infer information about the stratification and dynamics of different atmospheric 
structures. 

\noindent Wave propagation is an efficient means of carrying energy 
between different layers of the atmosphere and of dissipating it efficiently
through the formation and breaking of shocks \citep{mihalas}. Although acoustic (slow-mode)
heating is not important for the upper atmosphere (Athay \& White, 1979; White \& Athay, 1979), 
the role that wave propagation plays in the heating problem is still
one of the most challenging open debates amongst solar physics researchers.

The first measurements of oscillations in the quiet Sun 
\cite{leighton} came nearly a decade before the first 
detection of sunspot oscillations were reported \citep{beckers}.
Since then, many works have tried to put the pieces of the puzzle together
\cite[see][for a comprehensive review
of the literature of wave propagation in sunspots]{lites1992, bogdan2006}.

\noindent Oscillations above sunspots show a 5-minute periodicity in the photosphere.
However, at chromospheric levels and higher up, this picture changes into a 3-minute
saw-tooth pattern \citep[see e.g.][]{lites1986, lites1986b, sn2000,collados2001, bry2003, bry2004}.
Several hypotheses have attempted to explain the origin of the chromospheric 3-min
oscillations. Zhugzhda, Locans \& Staude (1985) proposed a model with a resonant 
chromospheric cavity that would explain the presence of multiple peaks in the
chromospheric power spectrum together with the  3 minute period. On the other hand, Gurman
\& Leichbacher (1984) suggested an origin based on the non-linear interaction of photospheric modes. 
Centeno, Collados and Trujillo Bueno (2006a) showed that the chromospheric 3-minute signal in the umbrae of sunspots is a result
of linear wave propagation of the photospheric perturbations in the 6 mHz range,
thus ruling out the non-linear origin for these oscillations.

Observations above facular and network areas report 5-minute oscillations
at chromospheric heights \citep[see e.g.][]{krijger, lites93, depontieu2003a, centeno2006a}
and higher up. For this to happen, the atmospheric
cut-off frequency must be reduced, allowing the "evanescent" photospheric 
5-minute oscillations propagate into the chromosphere. De Pontieu et al. (2003, 2004)
suggested that the inclination of magnetic fields plays an important role in
this p-mode leakage with enough energy to give rise to the dynamic
jets that are observed in active region fibrils. With this idea in mind,
Jefferies et al. (2006) suggested that inclined flux tubes might explain
the observed properties of waves at chromospheric heights.

The effective cut-off frequency is lowered by the cosine
of the inclination angle with respect to the local vertical. Thus, if the magnetic
field is sufficiently inclined, the flux tubes will channel the photospheric 
5-minute perturbations all the way up into the chromosphere and corona. An alternate 
possibility takes into account a departure from adiabaticity due to radiative
losses, which results in a reduced cut-off frequency \cite{roberts, centeno2006a, khomenko2008} that allows the 5-minute modes to propagate.

Simultaneous time-series
observations of various spectral lines that sample different regions of the 
solar atmosphere is one of the most useful techniques for studying wave 
propagation. 
For the analysis that follows we measure simultaneously the full Stokes vector of the 
photospheric Si {\sc i} \hbox{10827 \AA} line and of the chromospheric 
He {\sc i} \hbox{10830 \AA} multiplet on different magnetic targets. The analysis of the photospheric and 
chromospheric LOS velocity oscillations and the relation between them give 
us information about the behavior of the atmosphere: the propagation of 
photospheric disturbances, the amplification of the oscillations as they 
travel towards higher layers of the atmosphere, the cut-off frequency below which the 
oscillation modes do not propagate, the development of shocks and so on. 
We carry out a comparative study among 4 magnetic structures with 
different magnetic fluxes: a rather big sunspot (analyzed in Centeno et al. 2006a,
hereafter Paper I), a smaller one, a pore that has developed no penumbra and a 
facular region.
The physical properties of the atmosphere in each case determines how the 
propagation of the photospheric perturbations takes place and viceversa.

The outline of the paper is as follows: after a brief description of the observations
and the inversion procedures in Section 2, Section 3 continues to describe the properties
of the oscillations measured at photospheric and chromospheric heights.
Section 4 analyzes the relations between them, which are determined by the propagation
properties of the atmosphere in each case. The comparison with a simple model of linear vertical 
wave propagation in a magnetized atmosphere with radiative losses will yield to a
qualitative understanding of the origin of the chromospheric oscillations in the
analyzed magnetic structures. 
A brief discussion of the results is presented in Section 5.

\section{Observations}

\clearpage
\begin{figure}[t!]
\center 
\includegraphics[width=5cm]{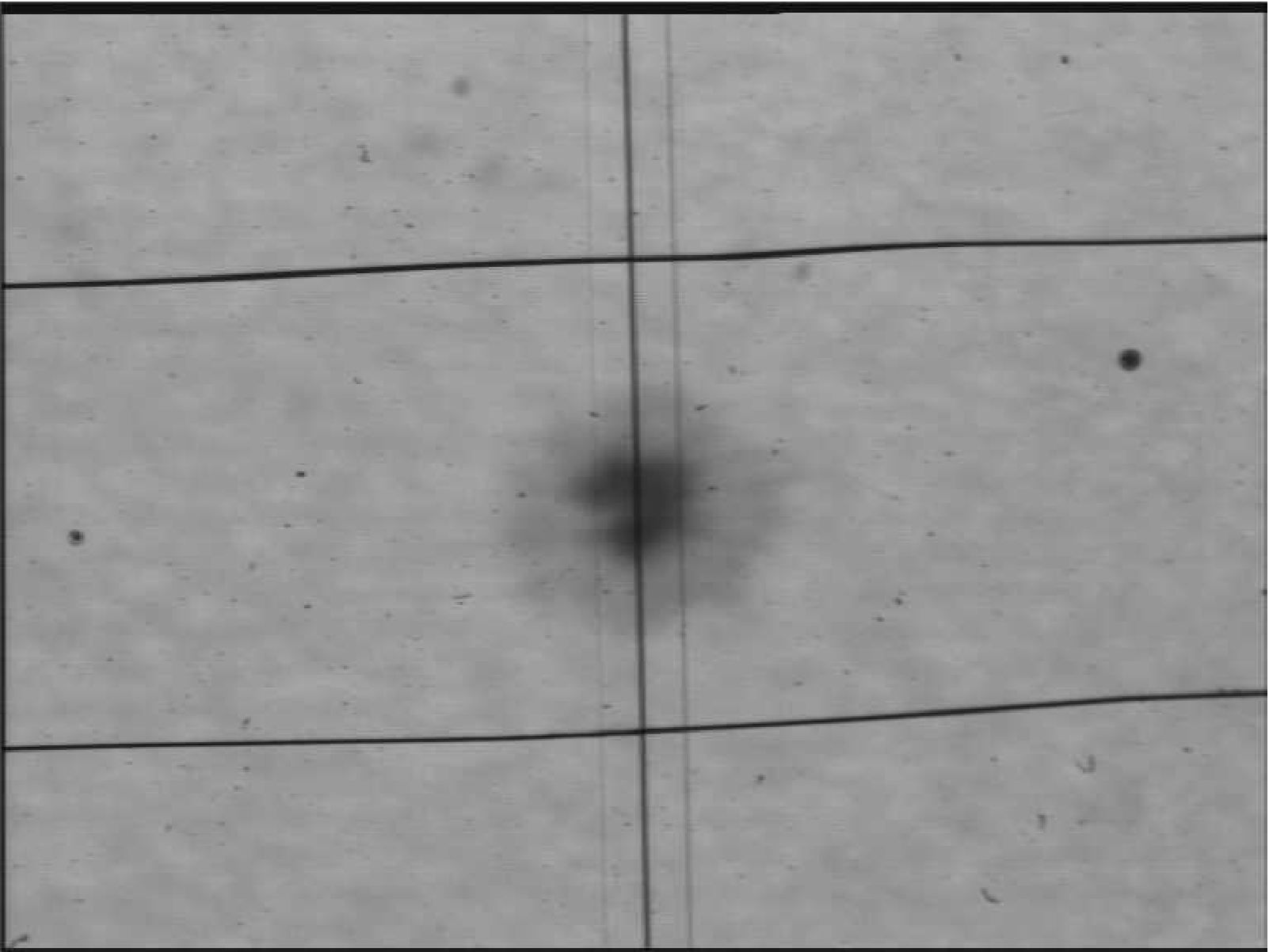} 
\includegraphics[width=5cm]{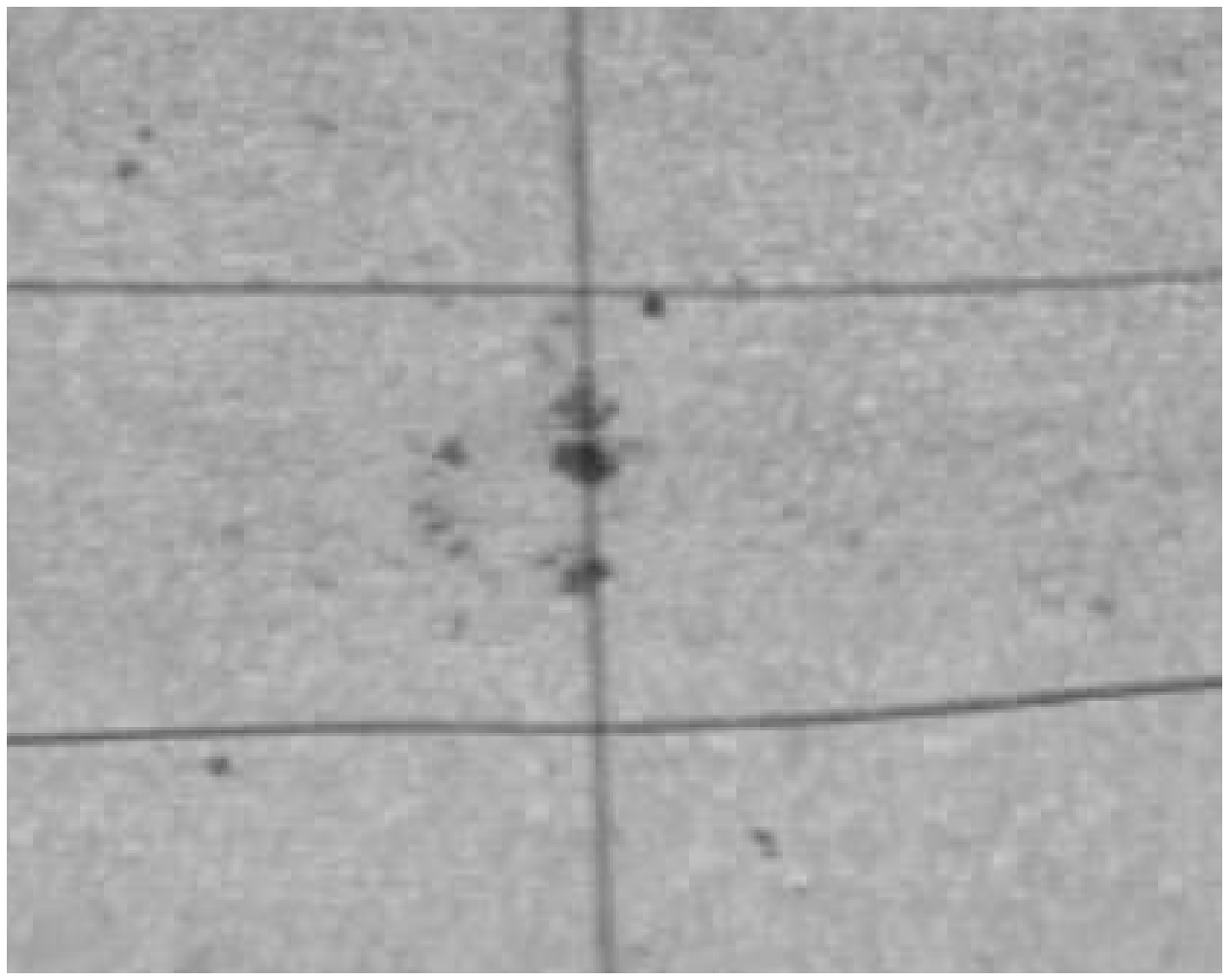} 
\includegraphics[width=5cm]{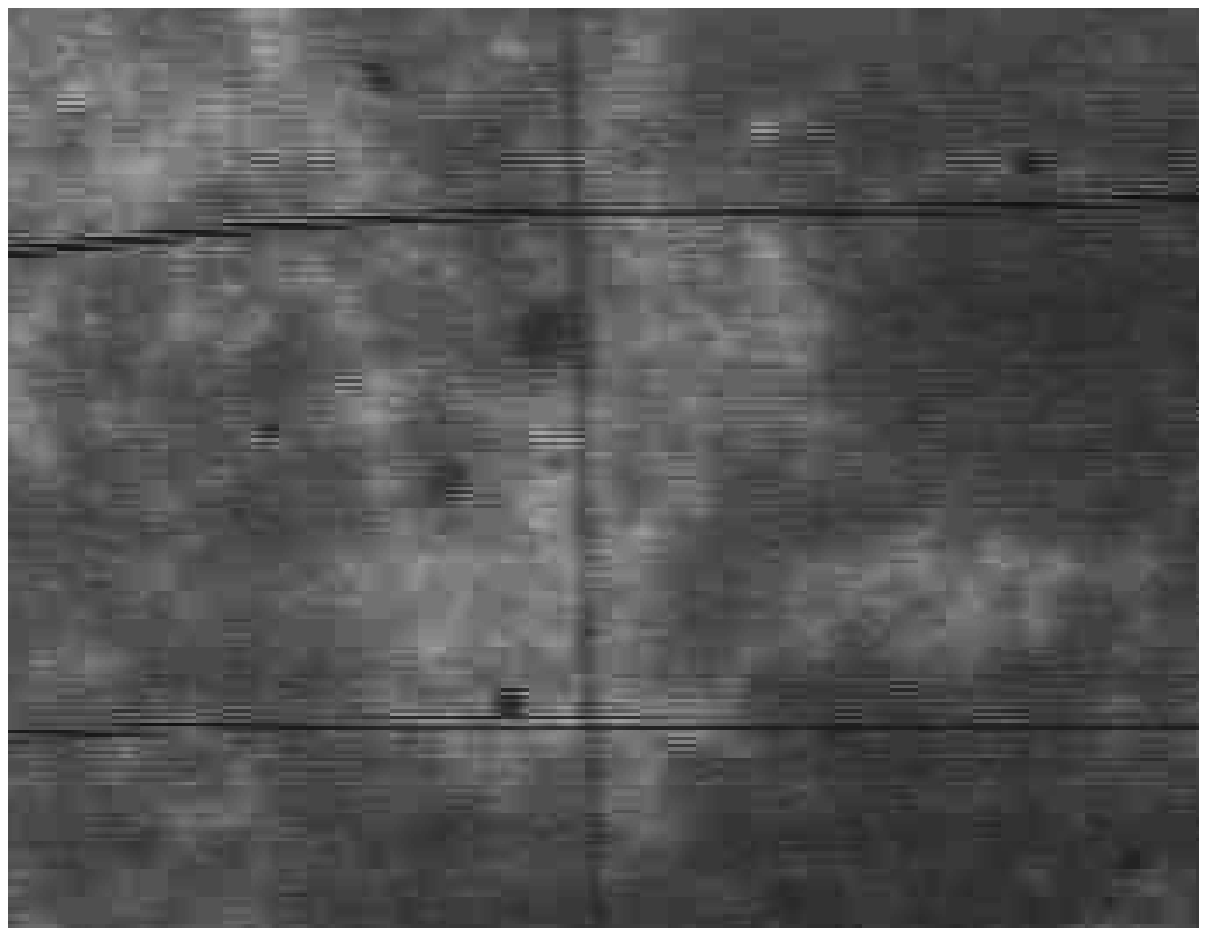} 
\caption{White light slitjaw images for the small sunspot and the pore, and Ca K image for the
facular region. In all the cases, the vertical line corresponds to the shadow of the spectrograph
slit (which was kept fixed during each run) and horizontal lines delimit the field of
view of the polarimeter ($\sim 40 \arcsec$). Of the three pores showed in the second image, 
we chose the central one to carry out the analysis presented in this paper.}
\label{fig:slitjaws}
\end{figure}
\clearpage

The observations were carried out at the German Vacuum Tower Telescope (VTT) 
of the Observatorio del Teide between September 2000 and June 2002, using the 
Tenerife Infrared Polarimeter \cite[TIP,][]{vmp-tip}. 
This instrument allowed us to measure simultaneously the four Stokes profiles for
all the spatial positions along the spectrograph slit at each scanning step.
The slit (0\farcs5 wide 
and 40\arcsec\ long) was placed over the targets and was kept fixed during 
the entire observing run (approx. 1 hour for each target). The image stability throughout 
the time series was achieved by using a correlation tracker device
 (Schmidt \& Kentischer 1995; Ballesteros et al. 1996)
which compensated, to first order, for the solar rotation as well as for the image motion 
induced by the Earth's high frequency atmospheric variability, thus minimizing the image jitter. Despite of being observed at very different moments, the quality is similar throughout the 
four datasets, and the instrument components and configuration were identical in all cases. 
The spatial resolution varies roughly between $1\arcsec$ and $1\arcsec.5$ (the worst data-set
being the one that corresponds to the smalls sunspot).
Details on the 4 data-sets analyzed in this paper are given in Table \ref{tab:data}. Fig. 
\ref{fig:slitjaws} shows the slit-jaw images for the small sunspot, the pore and the facula.
The size of the structure refers to the umbra in the case of the sunspots and to the 
enhanced bright region as seen in the Ca K slit-jaw in the case of the facula.
The data-set corresponding to the big sunspot was one of those used throughout Paper I.
We chose the more regular and homogeneous of the two sunspots and we only included it here for 
a sake of completeness, to compare it with the results from the other 3 magnetic structures.

Flat-field and dark current measurements were performed at
the beginning and the end of all observing runs, and, in order to
compensate for the telescope instrumental polarization, we also
took a series of polarimetric calibration images. The calibration
optics \cite[see][]{collados99} allows us to obtain the Mueller matrix of
the light path between the instrumental calibration subsystem
and the polarimeter.
This process leaves a section of the telescope
without being calibrated, so further corrections of the residual
cross-talk among Stokes parameters were done: the I to Q, U, and
V cross-talk were removed by forcing the continuum polarization
to zero, and the circular and linear polarization mutual cross-talks
were estimated by means of statistical techniques \cite{collados2003}.

\clearpage
\begin{table}[!t]
\begin{center}
\begin{tabular}{c|cccc}
 & big sunspot & small sunspot & pore & facular region\\
\hline
Date & May 9, 2001 & Sep 30, 2000 & Jun 13, 2002 & Jun 14, 2002 \\
Pos X [$\arcsec$] & -392 & 557 & -339 & -90 \\
Pos Y [$\arcsec$] & -163 & 102 & 306 & -291 \\
$\mu = {\rm cos} \theta$ & 0.89 & 0.81 & 0.88 & 0.94 \\
Duration [s] & 4214 & 3922 & 4842  & 4873\\
Cadence [s] & 2.1 & 7.9 & 5.4 & 5.4\\
Noise & $2 \cdot 10^{-3}$ & $7 \cdot 10^{-3}$&$2 \cdot 10^{-3}$ & $ 10^{-3}$\\
Size [$\arcsec$] & 16 & 10 & 4 & 30 \\
\label{tab:data}
\end{tabular}
\caption{Details of the four data-sets obtained with TIP. Positions X and Y represent terrestrial E-W and N-S directions and are measured from sun center.
}
\end{center}
\end{table}
\clearpage

The observed spectral range spanned from 10825 to 10833 \AA, with a 
spectral sampling of 31 m\AA\ per pixel. This spectral region is a powerful 
diagnostic window for the solar atmospheric properties since it contains 
valuable information coming from two different layers in the atmosphere. 
It includes a photospheric Si {\sc i} line at 10827 \AA\ and a chromospheric 
He {\sc i} triplet centered around 10830 \AA. 
The Si line is formed in the high photosphere. The response function (Ruiz Cobo \& del
Toro Iniesta, 1994) of the intensity profile to the temperature shows a height of formation
between 300 and 540 km above the base of the photosphere (Bard and Carlsson, 2008).
The He multiplet is formed in the high chromosphere \cite{avrett, knoelker, bruno07, centeno2008},
although the exact location depends critically on the atmospheric stratification and
the coronal illumination coming from above, that triggers the formation of the multiplet. 
Thus, the difference in height between the photospheric
and the chromospheric indicators ranges between 1000 and 1500 km. 
The He triplet serves as a unique diagnostic tool for 
chromospheric magnetic fields \cite[see][for a recent review]{lagg2007}.

In order to infer the physical parameters of the magnetized 
atmosphere in which the measured spectral lines were generated, we carried 
out the full Stokes inversion of both the silicon line and the helium triplet 
for the whole time series of observations and for all four data-sets. The Si
line was treated in Local Thermodynamic Equilibrium (LTE) and inverted with
the code LILIA (Socas-Navarro 2001). This inversion code yields the line-of-sight (LOS) 
velocity, magnetic field, temperature, density and electron
pressure stratification of the atmosphere in the layers where the spectral line
radiation is generated.
The observations of the He {\sc i} triplet were interpreted with our Milne-Eddington 
inversion code of Stokes profiles induced by the Zeeman effect, which is a suitable
strategy for extracting information on the LOS velocity and gives a reliable estimation
of the field strength from the full Stokes vector (see Trujillo Bueno \& Asensio Ramos 2007; 
Centeno et al, in press). We did not consider the incomplete Paschen-Back effect
in the modeling, so the magnetic field strength is underestimated by up to a 20\%
\cite{sasso}. However, in the analysis carried out in this paper we focus on the velocities,
which are not affected by this.
By inverting the whole time series we were able to obtain the temporal variability of several physical quantities (line-of-sight velocity, magnetic field intensity and orientation, 
\dots) at the photosphere and chromosphere of the four magnetic structures.
Both inversion codes took into account only one atmospheric component (one velocity and one
magnetic field value).
Only in the case of the facula, a stray light component was included to account for the 
non-magnetic part of the spectral profiles.

\noindent The magnetic field values yielded by the inversions are given in the line-of-sight (LOS)
reference system (which depends on the position of the target on the solar disk).
The azimuth origin, defined by the polarimetric calibration optics of the telescope,
is referenced to the Earth's North-South direction.
In order to determine how vertical the inferred magnetic fields are, we have to transform them
to the local vertical reference frame. The $180^{\circ}$ ambiguity in the azimuth 
leads to two possible inclination values in the new reference system. The flotability of
strong magnetic flux tubes can be used as a physical argument to choose the more vertical 
solution over the other option.
The photospheric magnetic fields obtained from the inversion of the Si line
turn out to be very vertical (with a range of inclinations between $0^{\circ}$ and 
$20^{\circ}$) in all the cases.
The He lines show barely any linear polarization at all due to the weaker magnetic field
regime in the high chromosphere.

\noindent Throughout the rest of this paper we will focus on the results concerning the 
LOS velocity oscillations, which are the line-of-sight projection of the plasma movements
along the magnetic field lines. 
All the targets are relatively close to disk center (the farther one having a heliocentric
angle of $\mu={\rm cos}\,\theta=0.81$), so this means that the maximum projection effect would happen for the
small sunspot, for which the LOS forms an angle of $\sim 35^{\circ}$ with the local vertical.
Taking into account the estimated height difference between the formation of the photospheric 
and the chromospheric indicators ($\sim1000 - 1500$ km), this angle would lead to a 
maximum projected horizontal displacement (for two positions on the same vertical) of some
$900$ km. The spatial resolution in our data was limited by seeing, which
we estimated to be of the order of $\sim1 - 1.5\arcsec$, so the displacement due to projection 
effects will be barely noticeble in the worst case.

\section{Oscillations and Shock Waves}

\clearpage
\begin{figure}[t!]
\center 
\includegraphics[width=7cm]{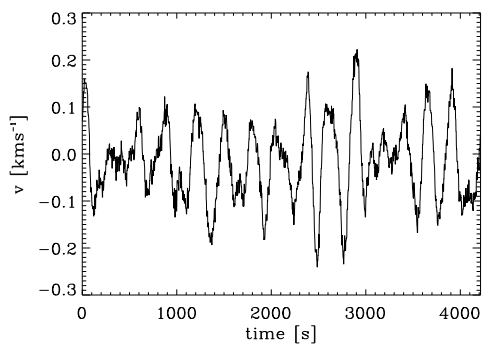} 
\caption{Photospheric oscillations for one position inside the umbra of
sunspot \#1. The rest of the structures analyzed in this work show a 
qualitatively similar behavior at photospheric levels.}
\label{fig:photosphere}
\end{figure}

\begin{figure}[t!]
\center 
\includegraphics[width=7cm]{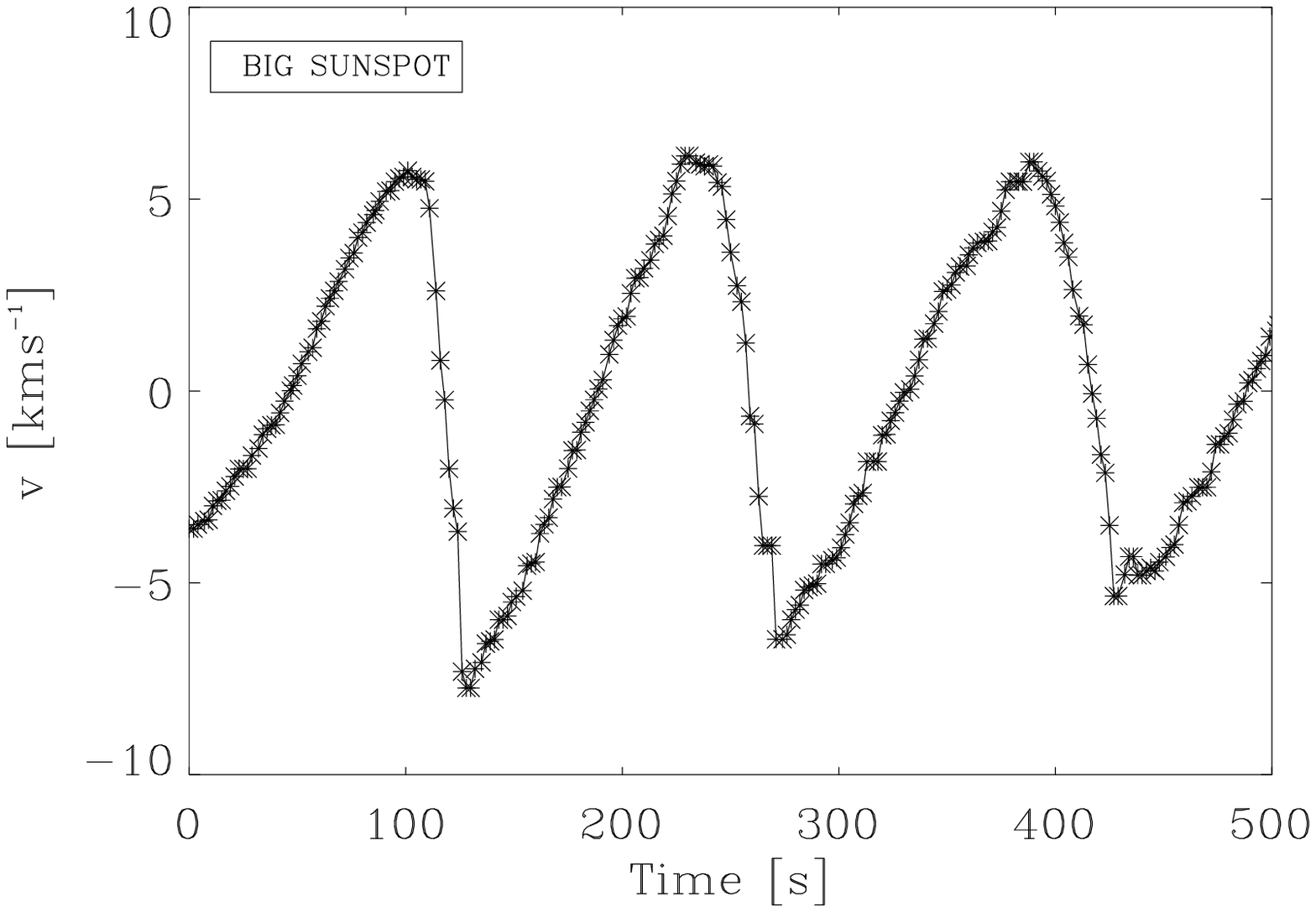} 
\includegraphics[width=7cm]{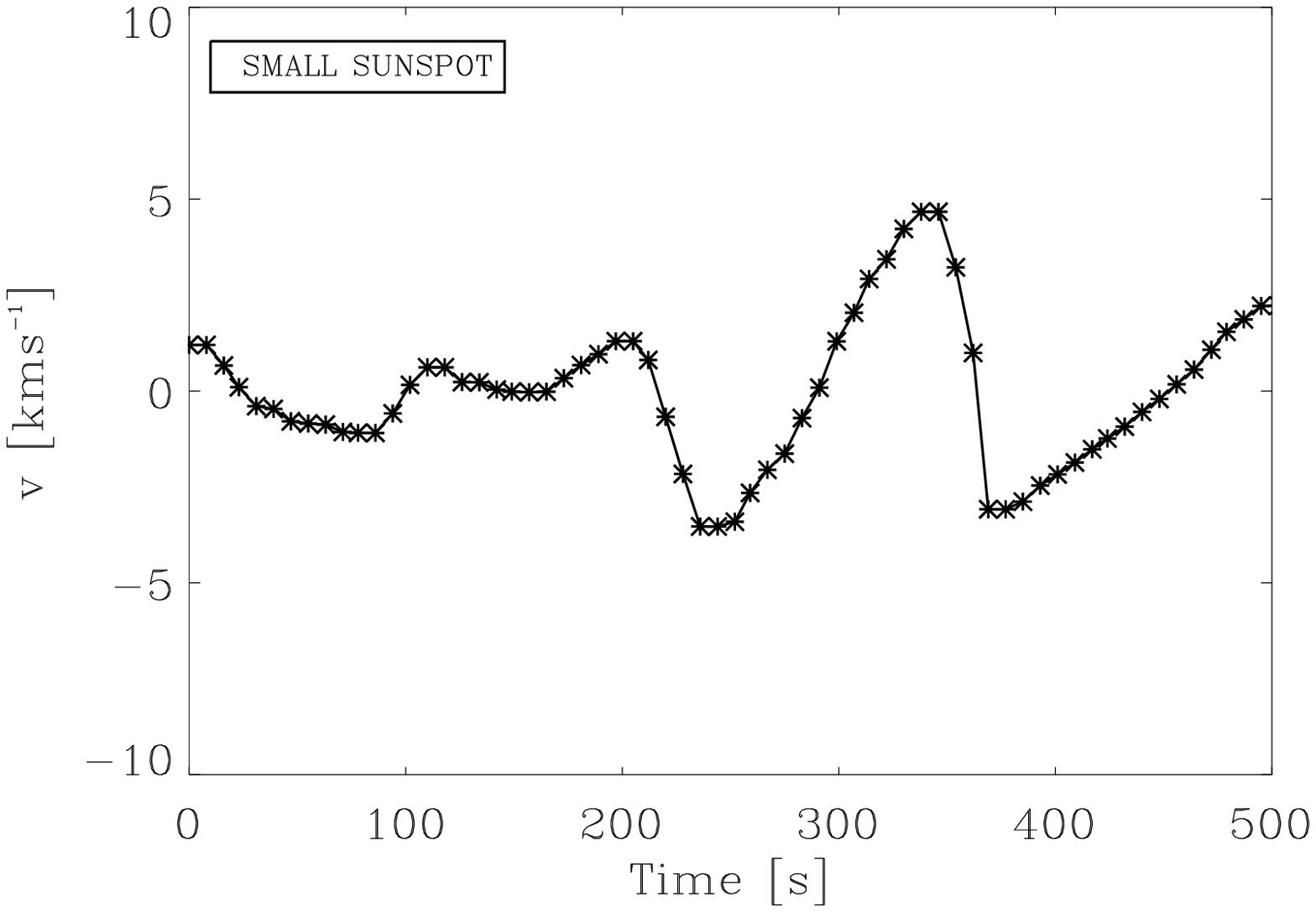} 
\includegraphics[width=7cm]{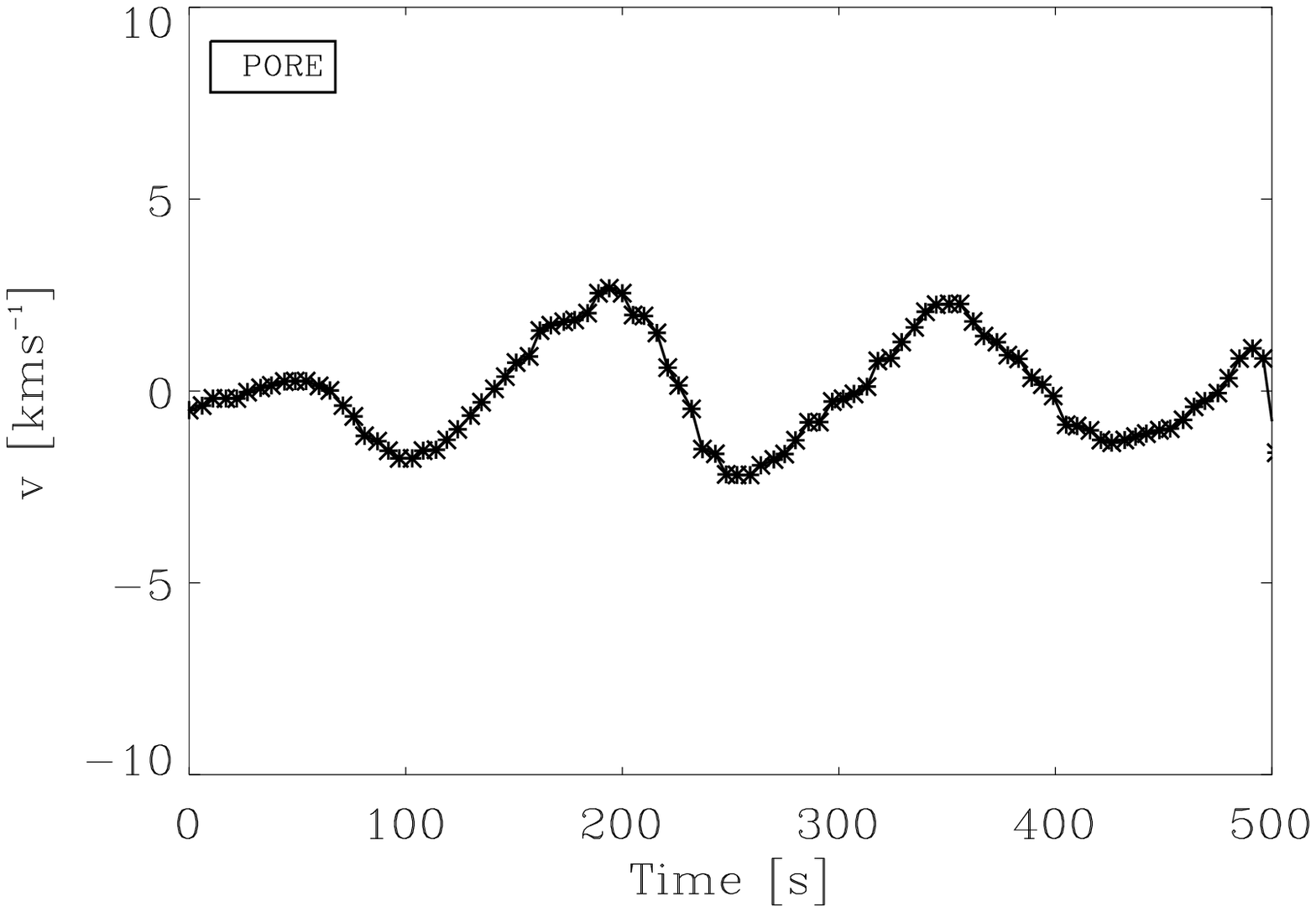} 
\includegraphics[width=7cm]{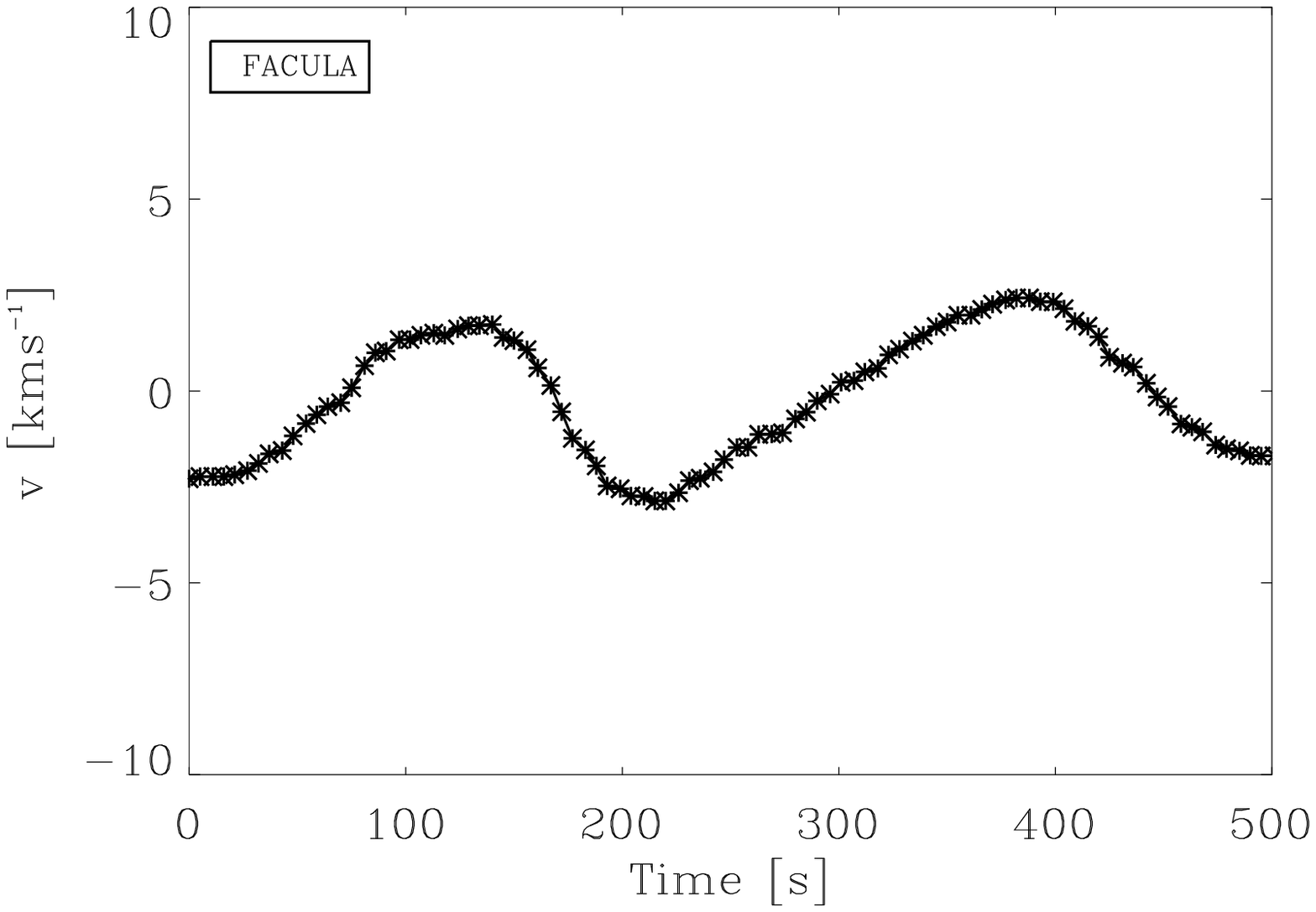} 
\caption{Chromospheric oscillations in different magnetic features as
seen by the He {\sc i} 10830 \AA\ triplet. 
From left to right and top to bottom, the panels depict a typical 
velocity profile for one position inside the umbra of a big sunspot, the 
umbra of small sunspot, a pore and a facula. In all the cases, 
asterisks represent the measured values, equispaced in time.}
\label{fig:chrom-oscil}
\end{figure}
\clearpage

Photospheric velocity\footnote{The sign convention used throughout this 
paper is such that negative velocities correspond to material approaching 
the observer along the line-of-sight.} oscillations 
(see Fig.~\ref{fig:photosphere}), 
retrieved from the inversion of the Si {\sc i} line, show the same 
characteristics everywhere: a typical 5-min period, 300--400 m\,s$^{-1}$ 
peak to peak amplitudes and fairly sinusoidal patterns. 

\noindent On the other hand, chromospheric velocity oscillations (encoded in
the Doppler shift of the He {\sc i} triplet) show very different
behavior depending on the magnetic structure. Fig.~\ref{fig:chrom-oscil} 
depicts a detail of the chromospheric oscillation pattern in the four
regions analyzed. From left to right and top to bottom, the
panels show the velocity variations in the umbrae of the big and medium-sized sunspots,
the pore and the facular region. The asterisks represent the measured 
values, equispaced in time. The departure from a sinusoidal behavior
is a signature of the passage of shock waves through this 
layer of the atmosphere.
It is clear from Fig.~\ref{fig:chrom-oscil} that, as the magnetic flux
of the structure decreases, the amplitude of the oscillations also becomes
smaller (from $\sim$15 km\,s$^{-1}$ in the big sunspot to $\sim$3--4 km\,
s$^{-1}$ in the facula). Note that the projection effects are larger for the three
sunspot-like features than for the facula because they are farther away from disk center. 
Assuming that the plasma movements are directed along the field lines - parallel to the
local vertical- then the actual oscillation amplitudes for the sunspots are even larger than
the values given above.
The steepness and the frequency of appearance of the shocks is also correlated with
the magnetic flux of the observed feature. 

\noindent The chromospheric oscillations in the facular region present a 
particularity that is not shared by the sunspot-like structures (i.e. both
sunspots and the pore). While the latter show a characteristic 3-minute pattern, 
the facular region presents, in contrast, an obvious 5-minute period at 
chromospheric heights.

\clearpage
\begin{figure}[t!]
\center 
\includegraphics[width=7cm]{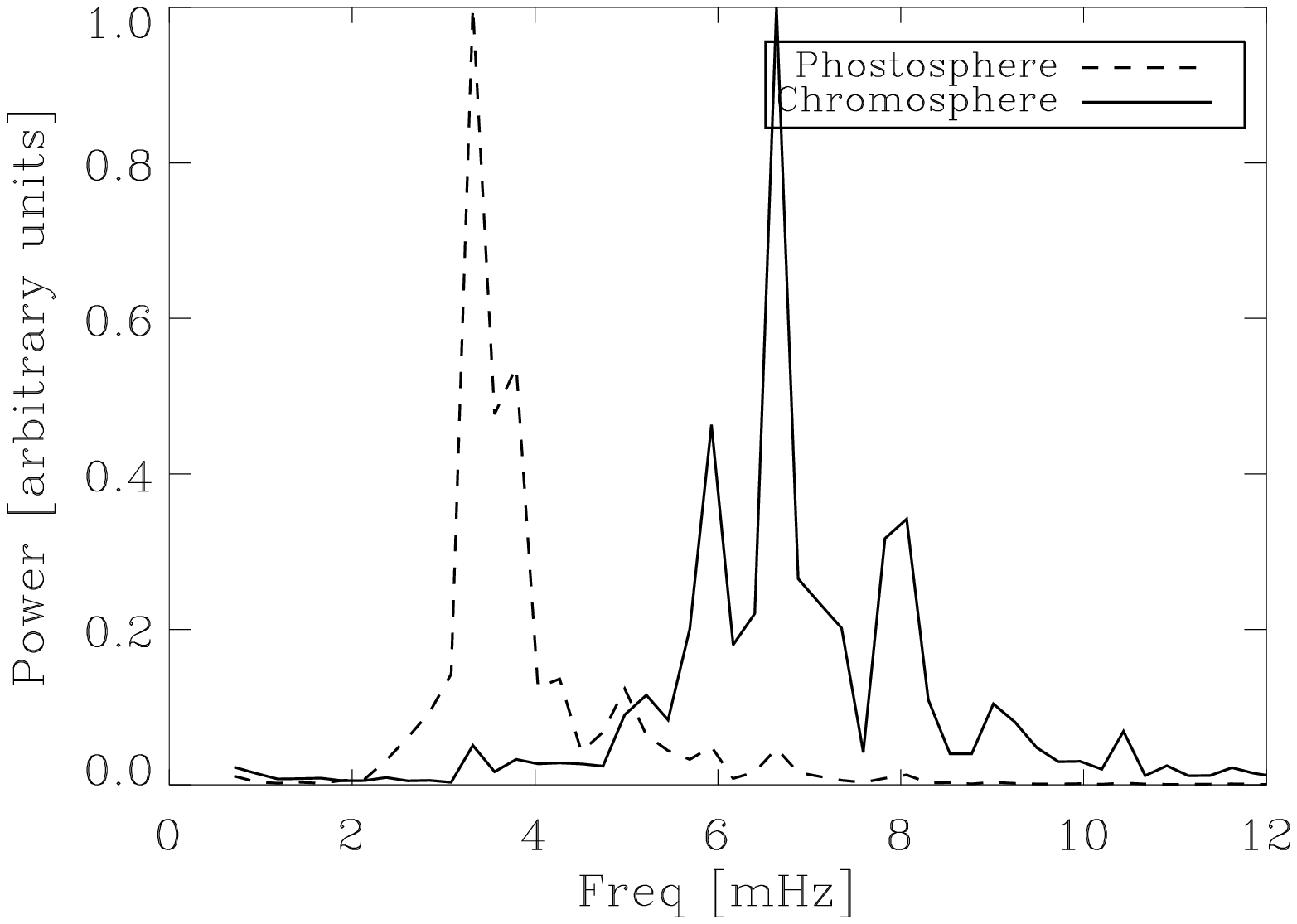} 
\includegraphics[width=7cm]{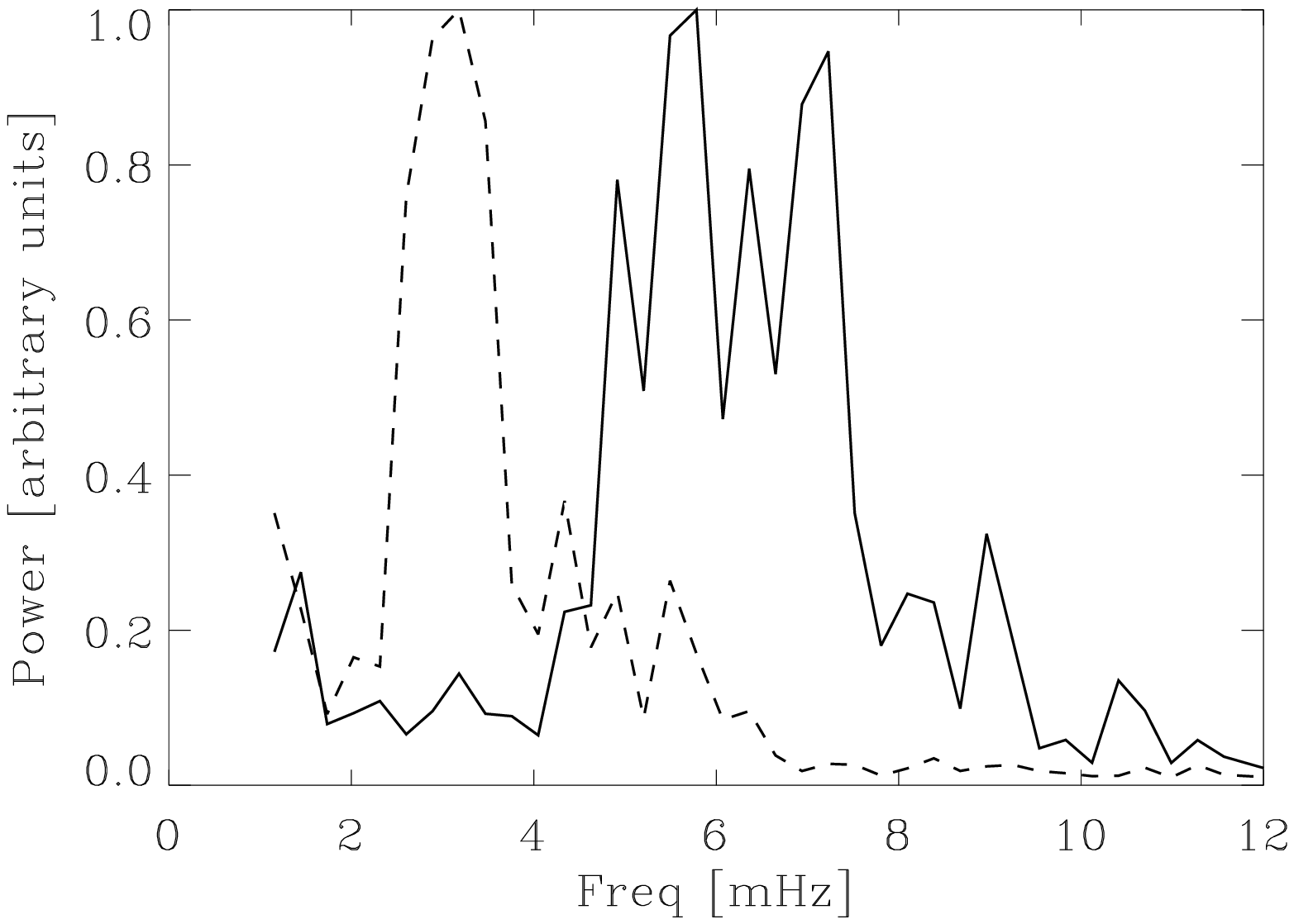} 
\includegraphics[width=7cm]{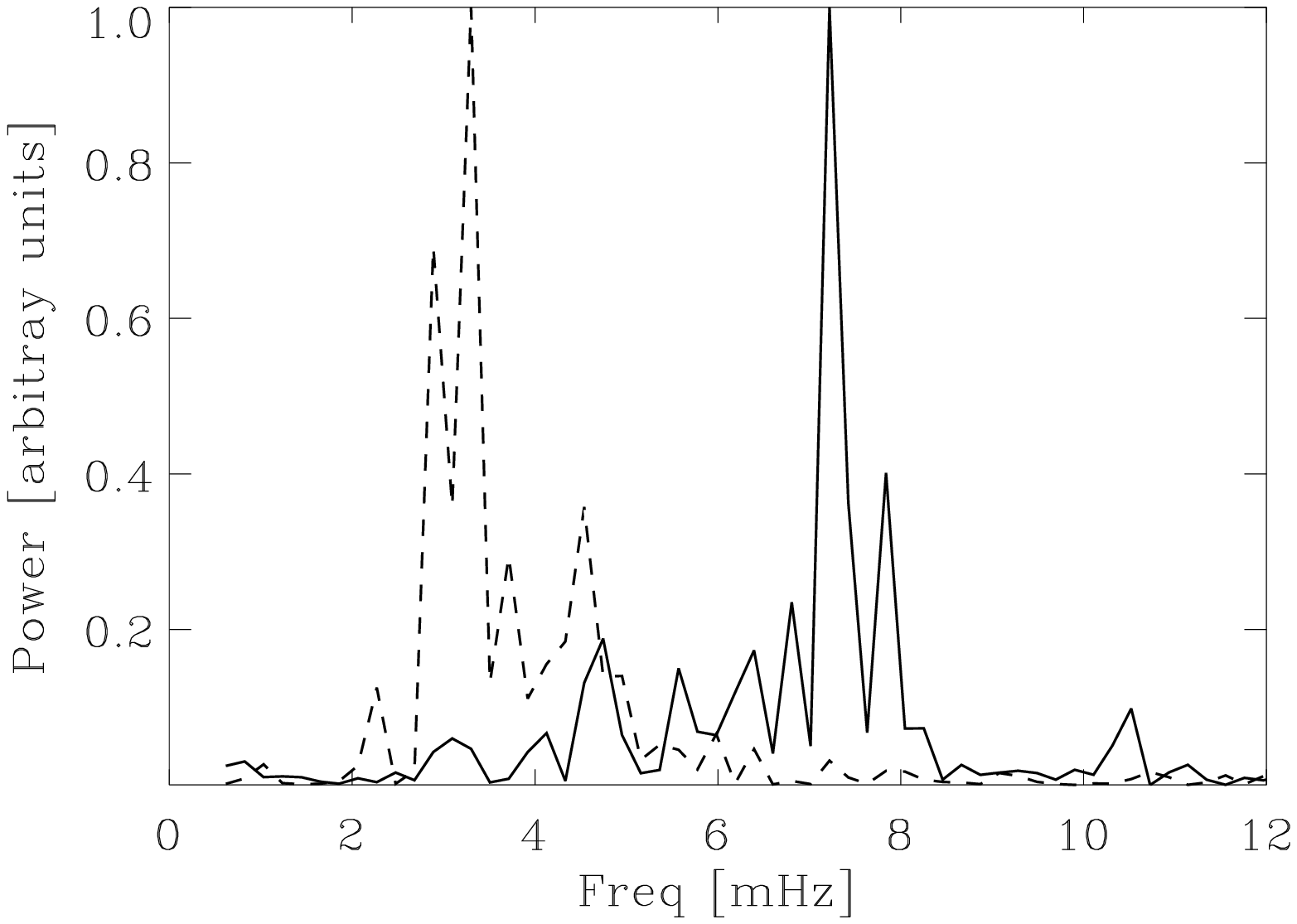} 
\includegraphics[width=7cm]{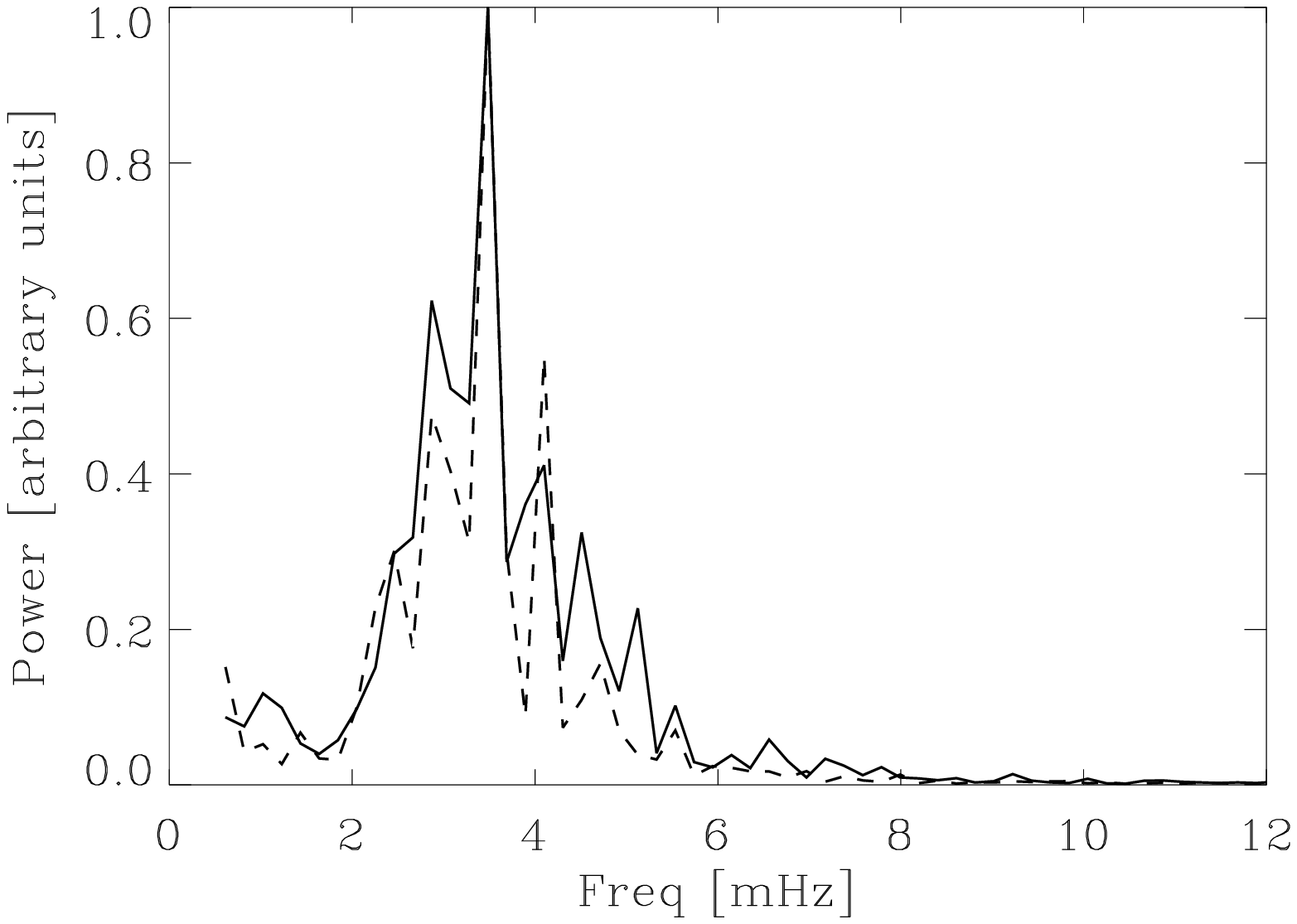} 
\caption{Normalized LOS velocity power spectra of the four analyzed regions. 
From left to right
and top to bottom, the panels show the photospheric (dashed) and chromospheric
(solid) power spectra, (averaged over the length that the structure spanned
along the spectrograph slit and normalized to their maximum), for the big sunspot, the small sunspot, the pore
and the facular region.}
\label{fig:power-spectra}
\end{figure}
\clearpage

Fig.~\ref{fig:power-spectra} illustrates the difference between
the spectral composition of the oscillation patterns of the sunspot-like 
structures and the facular region.
From left to right and top to bottom, the panels represent the 
average photospheric ({\em dashed}) and chromospheric ({\em solid}) power
spectra in the four magnetic features. While in the sunspot-like
structures the majority of the oscillatory energy is concentrated around 
different frequency regimes in the photosphere and the chromosphere, the 
facular power spectra show a strikingly similar behavior at both heights 
(with the maximum power lying at 3.3\,mHz and the secondary peaks being 
co-located and showing pretty much the same distribution).

\noindent One could attribute this to the helium triplet being formed lower in the atmosphere
(closer to the region of formation of the Silicon line).
However, this can be ruled out for two reasons: 1) the oscillations derived from
the He lines have at least ten times the amplitude of those measured with the Si line,
indicating that they are clearly chromospheric. 2) if the He triplet is formed
by the triggering effect of the coronal irradiance \cite[see][]{avrett, centeno2008}, there is no way the lines can have a contribution function from the 
photosphere, but it can only come from the high chromosphere.

\noindent The 5-minute chromospheric oscillations appear as a consequence of the reduced
cut-off frequency in facular regions.

\section{Wave propagation}

For the wave propagation analysis we will follow the strategy adopted in
Paper I.
First, we will analyze the information given by the mean power, phase difference and amplification
spectra. This will allow us to identify the propagation regime of the wave modes in each case
and determine the cutoff frequency and the amplification of the oscillations as the waves travel
through the atmosphere. Then, we will try to reproduce these observables with a simple model 
of vertical linear wave propagation along constant magnetic field lines in a stratified 
atmosphere. The theoretical modeling will allow us to make a prediction about the time 
delay between the signals measured at the photosphere and the chromosphere, which will be
contrasted with what is obtained by mere cross-correlation of the measured velocity maps
at both heights.

\subsection{Phase difference and cutoff frequencies}

\clearpage
\begin{figure}[t!]
\center 
\includegraphics[width=7cm]{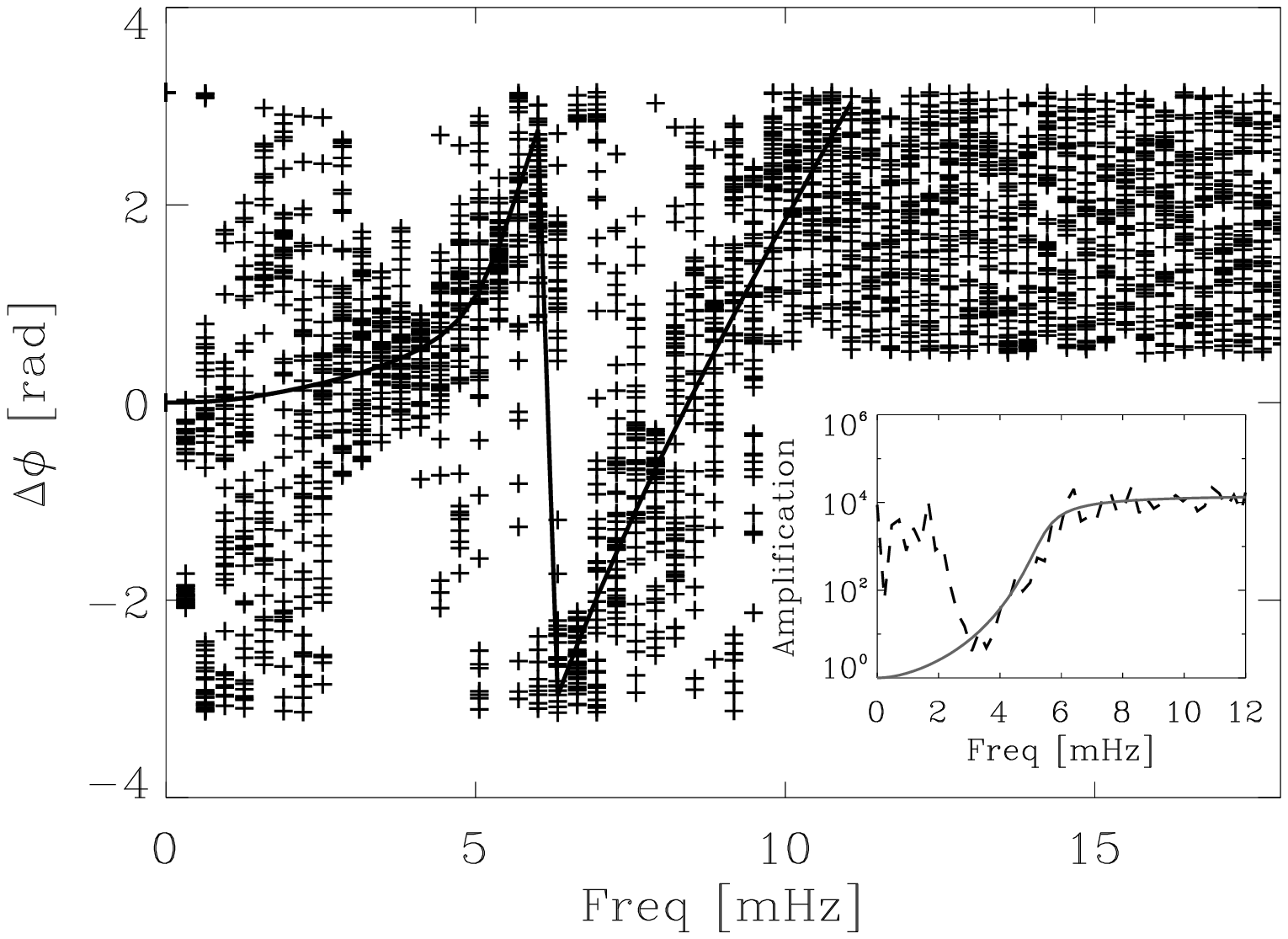} 
\includegraphics[width=7cm]{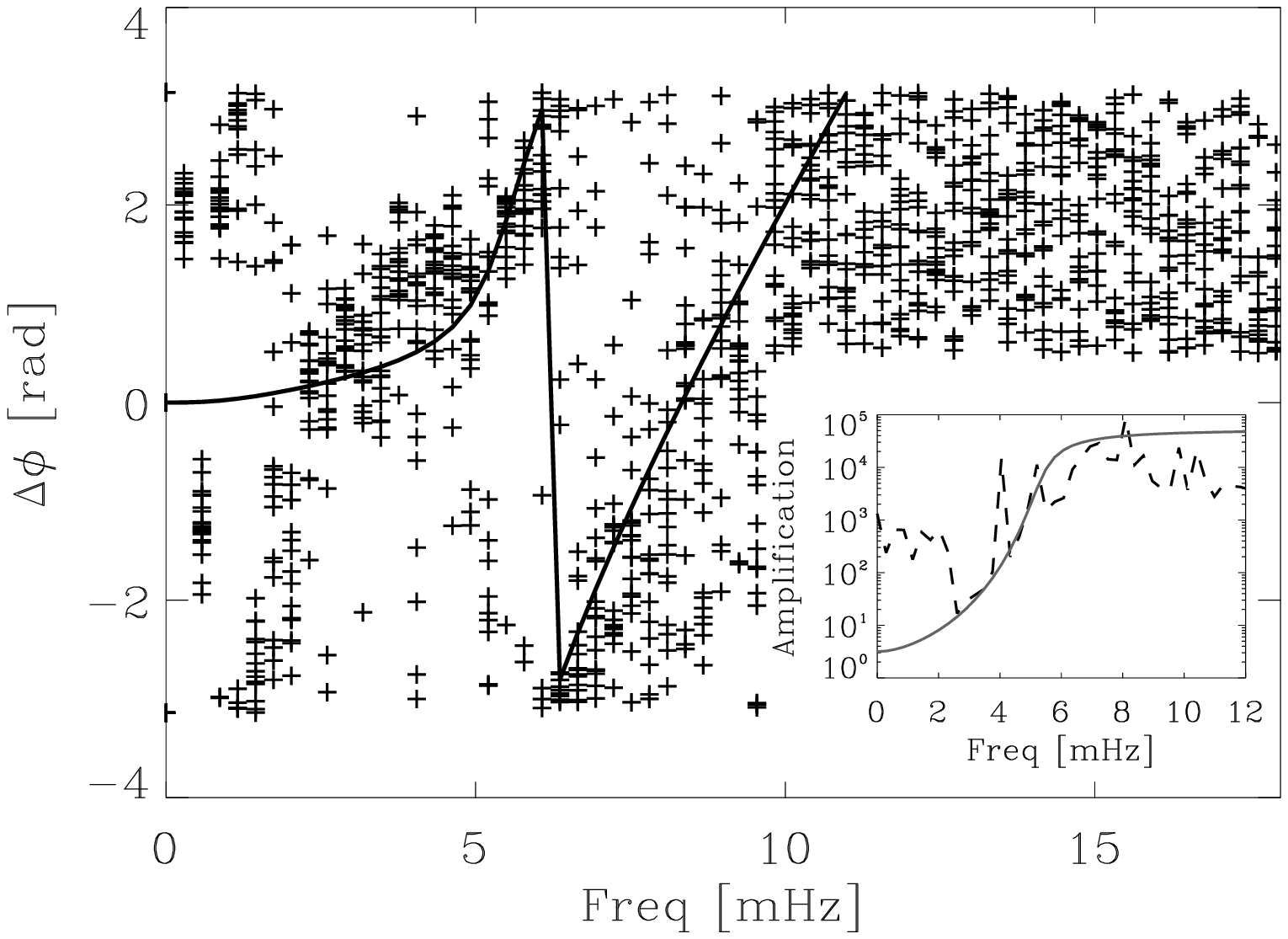} 
\includegraphics[width=7cm]{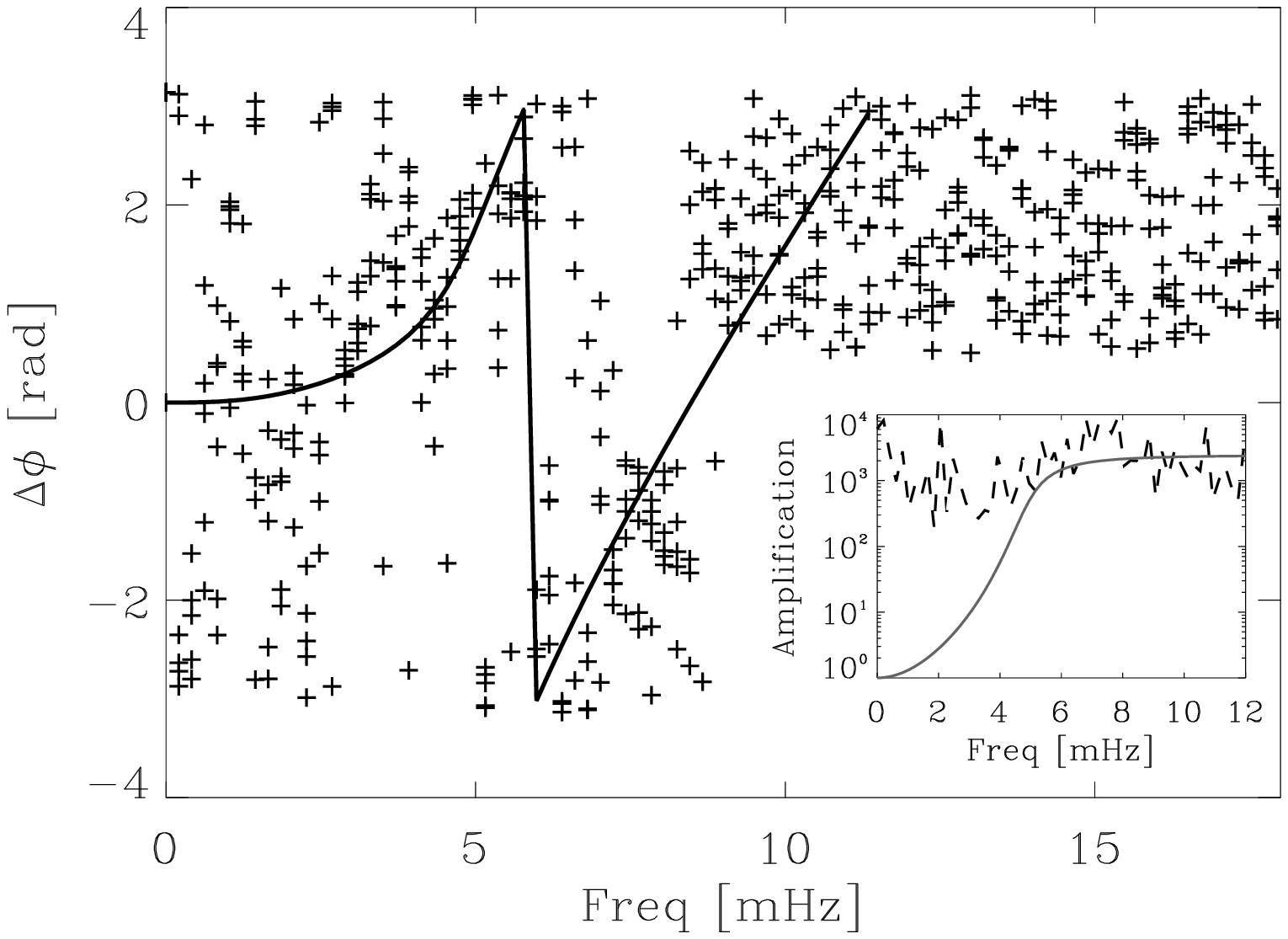} 
\includegraphics[width=7cm]{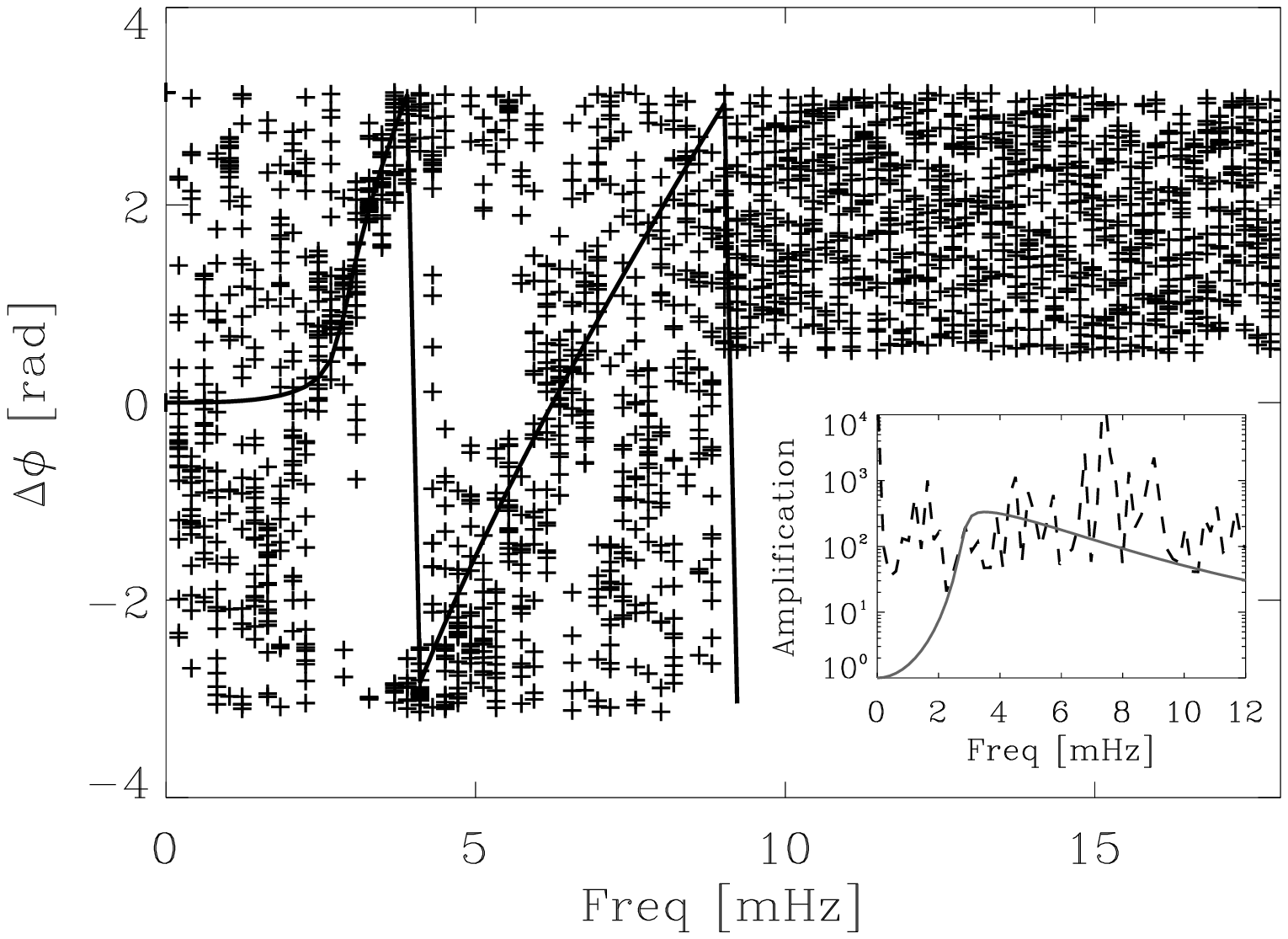} 
\caption{Phase difference between photospheric and chromospheric oscillations 
as a function of frequency for the four magnetic structures (from left to right and top to bottom
the panels correspond to the umbrae of the big and small sunspots, the pore and the facula). Each cross on 
the figure was computed as the difference between the phase of the 
chromospheric and the photospheric oscillation for a unique position along
the slit and a particular frequency. The small insets show the respective
amplification spectra. Solid lines correspond to the best fit to a model of
linear wave propagation in a stratified isothermal atmosphere with radiative losses,
permeated by a constant vertical magnetic field.}
\label{fig:phase}
\end{figure}
\clearpage

Fig.~\ref{fig:phase} shows the 
phase difference between the photospheric and the chromospheric 
oscillations for the four magnetic features analyzed in this paper.
Each cross on the panels was computed as the difference between the phase of 
the chromospheric and the photospheric oscillations for one position along
the slit and one frequency value.
The turn-off points in the phase spectra 
correspond to the effective cut-off frequencies of the 
atmosphere. Oscillations with lower frequencies will not be able to make it 
through the atmosphere and, thus, will remain trapped producing
stationary waves. Above the cut-off value, perturbations propagate upwards 
freely into the chromosphere.
Note that in the case of the sunspot-like structures the atmospheric 
cut-off stands around 4 mHz, inhibiting the propagation of lower
frequency modes. This has strong implications on what happens
to the photospheric oscillatory energy -which lies
below the cut-off value- and strikes the question of the origin of the 
3-minute chromospheric power in such structures. Paper I addressed this question 
and proved that, inside the umbrae of sunspots, the chromospheric 
3-minute oscillations come from the upward linear propagation of the photospheric
3-minute power rather than from the non-linear interaction and 
redistribution of the energy stored in the 5-minute photospheric modes.
This means that, while the 5-minute photospheric component stays trapped in the
atmosphere giving rise to stationary waves, the 3-minute photospheric
component travels upward through the atmosphere driving the chromospheric 
oscillations.

\noindent The case of the facular region is somewhat different. Now, the
cut-off frequency is lower than that of the typical photospheric modes, 
thus allowing them to propagate through the atmosphere.
This would explain why the facular chromospheric power spectrum peaks around
3.3 mHz, since all the 5-minute photospheric power can travel freely upward
into the high chromosphere. 

\noindent The small inset in the lower right corner of each panel of
Fig.~\ref{fig:phase} represents
the corresponding amplification spectrum (dashed line), this is, the ratio of the
chromospheric power over the photospheric power as a function of
frequency. For decreasing magnetic flux, the amplification of the power
in the propagation regime (i.e. above the cut-off frequency) also decreases.

\subsection{Model}

A simple model of upward linear wave propagation in an isothermal stratified 
atmosphere permeated by a constant vertical magnetic field (described in Paper I, 
but originally accounted for by Souffrin in 1972) was chosen to further explain the 
observations. Energy exchange by radiative losses is permitted by Newton's 
cooling law, which accounts for the damping of the temperature fluctuations
with a typical relaxation time $\tau_R$ \citep{spiegel57, kneer87}:

\begin{equation}
\tau_R=\frac{\rho c_v}{16 \chi \sigma_RT^3}
\end{equation}

\noindent where $\chi$ is the grey absorption coefficient and $\sigma_R$ the 
Stefan-Boltzmann constant. The solution $A(z) = D e^{z/2H_0} e^{ik_z z} e^{i\omega t}$ substituted
into the wave equation:

\begin{equation}
\hat c^2\frac{d^2 A(z)}{dz^2} -\hat \gamma g \frac{dA(z)}{dz} +\hat \omega^2 A(z) = 0
\end{equation}

\noindent yields a dispersion relation:

\begin{equation}
k_z^2 = \frac{\omega^2-\hat\omega_{ac}^2}{\hat c^2},
\end{equation}

\noindent where $k_z$ is the vertical wave number, $H_0$ the pressure scale height,
and $\hat\omega$ and $\hat c$ were defined in \cite{bunte}:

\begin{equation}
\hat \omega_{ac} = \hat c / 2H_0, \qquad \hat c^2 = \hat \gamma g H_0, \qquad \hat \gamma = \frac{1-\gamma i \omega \tau_R}{1 - i \omega \tau_R}
\end{equation}

\noindent This results in a 3-free-parameter model, with the 
temperature, $T$, the vertical distance between the measured oscillations, $\Delta z$, 
and the typical radiative cooling time, $\tau_R$, as the fitting coefficients.
For a more detailed description, we refer the reader to Paper I, B\"unte and Bogdan (1994), Mihalas \& Wiebel-Mihalas (1984) and Souffrin (1972).

\noindent The model assumes a magnetic field that is aligned with gravity. The full
Stokes inversion of the photospheric Si line in our data-sets shows that, in all cases,
the deviation of the magnetic field direction from the local vertical never exceeds 
$20^{\circ}$, being close to $0^{\circ}$ for most of the pixels (within the uncertainties).

The solid lines (both in the bigger panels and in the insets) of Fig. \ref{fig:phase}
correspond to the best fit of the model to the data. There, all the pixels of the observed 
targets are plotted together. The free parameters of the model were adjusted to explain, simultaneously, the average observed phase difference and 
average amplification spectrum. The fit has to account for the turn-off point 
due to the cut-off frequency and the steepness of the phase spectrum together 
with the magnitude of the amplification of the power. The different parameters have very
distinct effects on the resulting curves. For instance, while the radiative cooling time controls
the position of the turn-off point, the height difference determines the magnitude of the
amplification of the signals from the photosphere to the chromosphere. 
Table \ref{tab:fitting-params} compiles the values of the resulting 
fitting parameters.

\clearpage
\begin{table}[!t]
\begin{center}
\begin{tabular}{c|cccc}
 & big sunspot & small sunspot & pore & facular region\\
\hline
$T$ [K] & 4000 & 4500 & 5000 & 9000 \\
$\Delta z$ [km] & 1000 & 1000 & 1000 & 1500 \\
$\tau_R$ [s] & 55 & 30 & 25 & 10 
\label{tab:fitting-params}
\end{tabular}
\caption{Fitting parameters used in linear wave propagation modeling}
\end{center}
\end{table}
\clearpage

In the sunspot-like structures, as the magnetic flux decreases, the typical
cooling time also decreases while the temperature grows. Although quantitatively
these numbers are difficult to justify due to the simplicity of the model, 
they do make sense in a qualitative way.
It is no doubt expected that the smaller the structure, the larger the temperature 
inside it, since its magnetic field becomes more and more incapable of inhibiting the 
convection process underneath the photosphere. On the other hand, the radiative
cooling time is related to the inhomogeneity, taking smaller values the less homogeneous
the structure \citep[see][]{spiegel57, kneer87}. The difference in heights remains essentially the same for the three sunspots,
while it turns out to be larger in the case of the facula. Sunspot-like features are evacuated
structures (due to the balance between magnetic and gas pressure), allowing the coronal
EUV irradiance to travel further down into the chromosphere and trigger the formation of
the He {\sc i} 10830 \AA\ triplet at lower layers \cite[see][]{centeno2008}.

\subsection{Theoretical prediction and measured time delays}

\clearpage
\begin{figure}[t!]
\center 
\includegraphics[width=7cm]{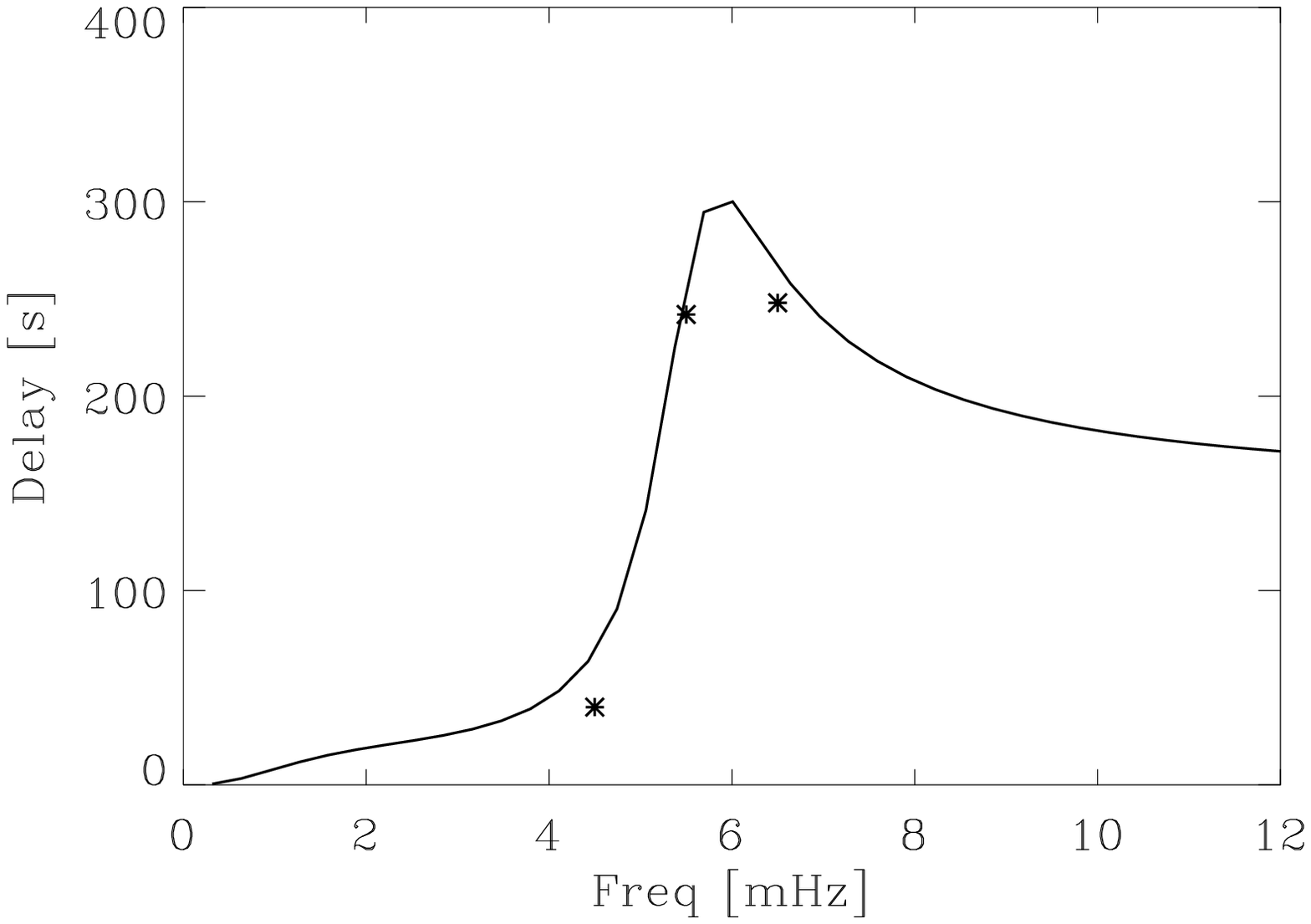} 
\includegraphics[width=7cm]{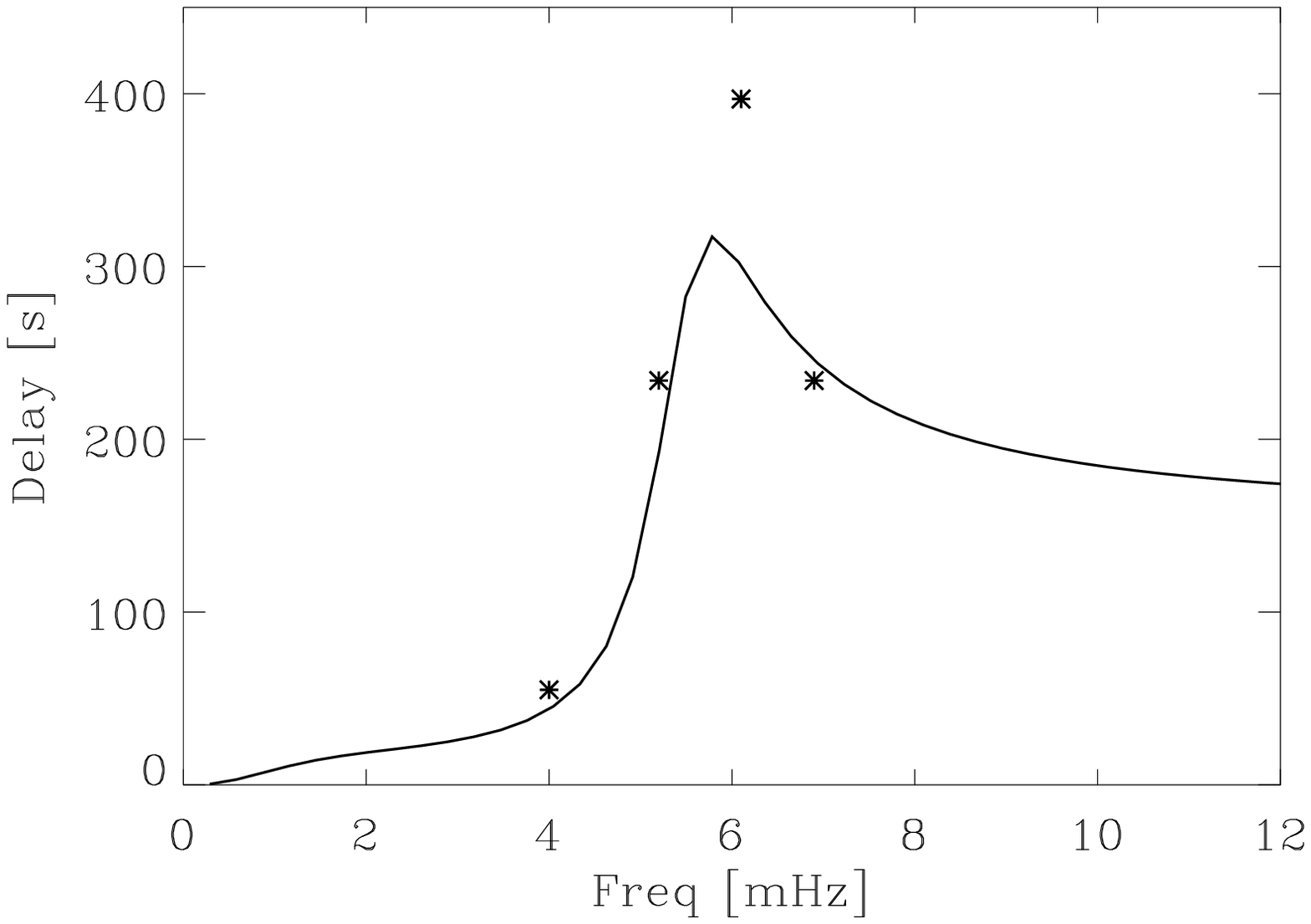} 
\includegraphics[width=7cm]{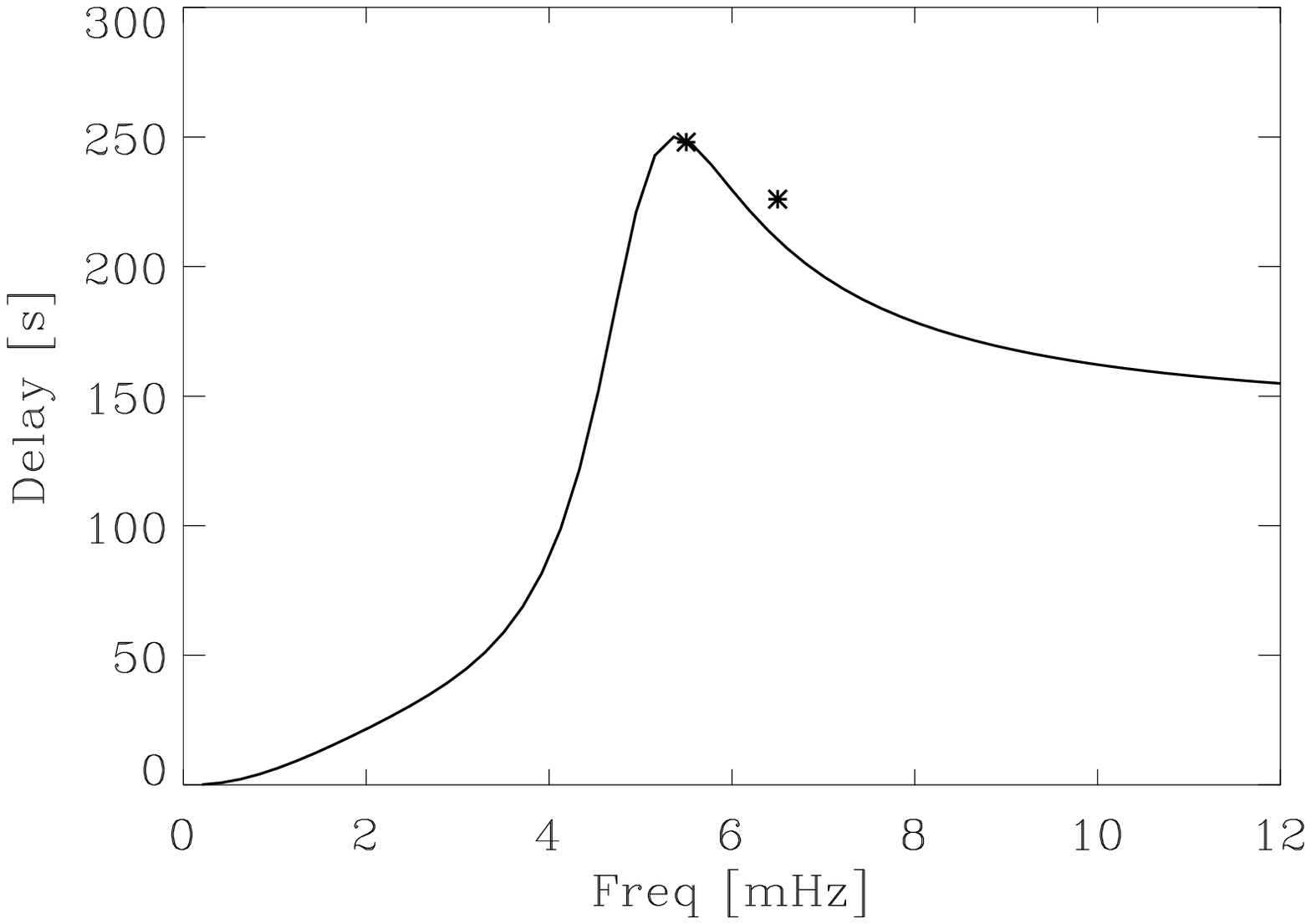} 
\includegraphics[width=7cm]{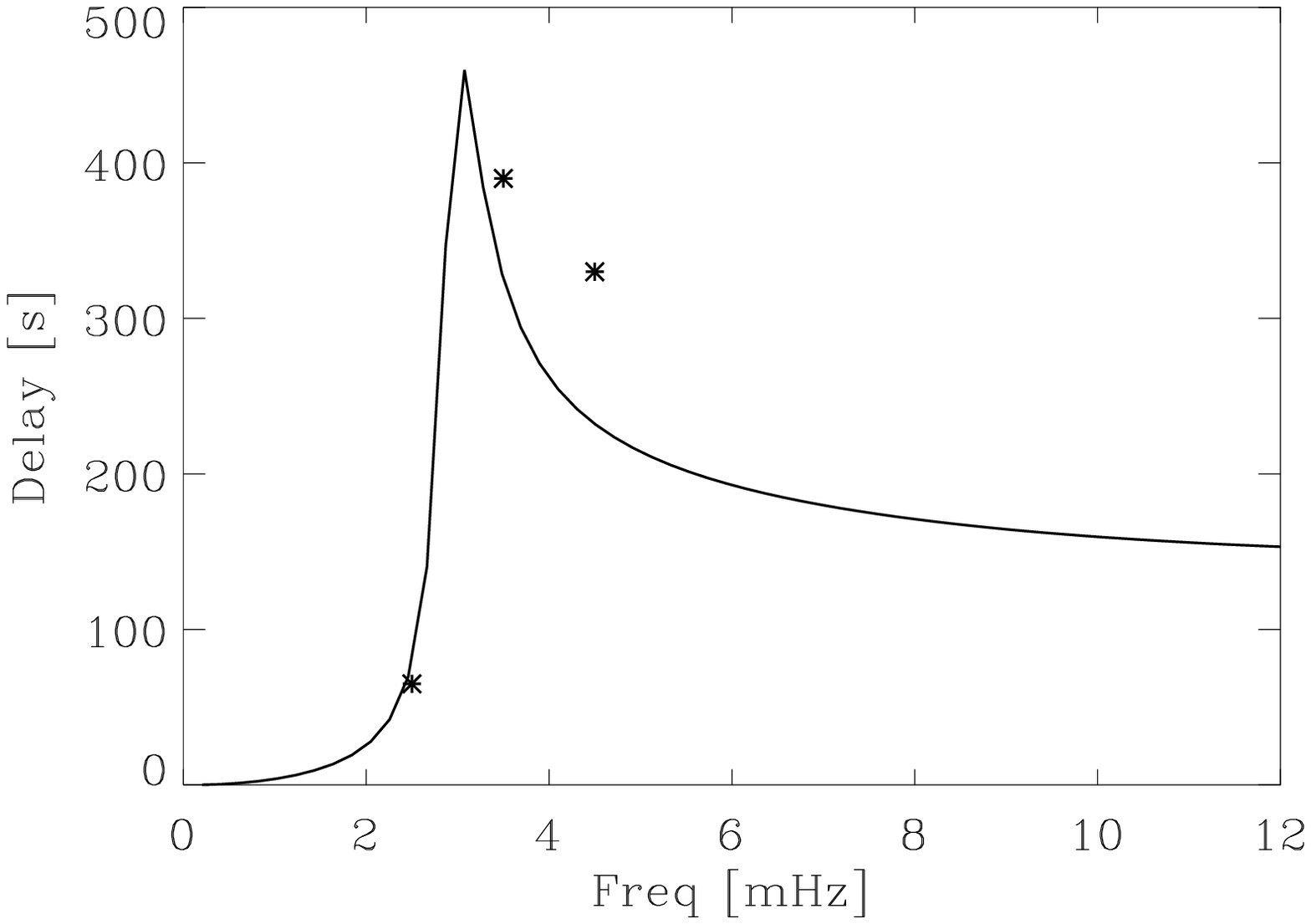} 
\caption{Expected and measured time delays. From left to right and top to bottom,
the panels correspond to the umbra of the big sunspot (extracted from Paper I), 
the umbra of the small sunspot, the pore and the facular region. The solid line shows 
the expected time delay, as a function  of frequency, predicted by the model. The asterisks correspond to the measured delays. }
\label{fig:theor-time-delay}
\end{figure}

\begin{figure}[t!]
\center 
\includegraphics[width=9cm]{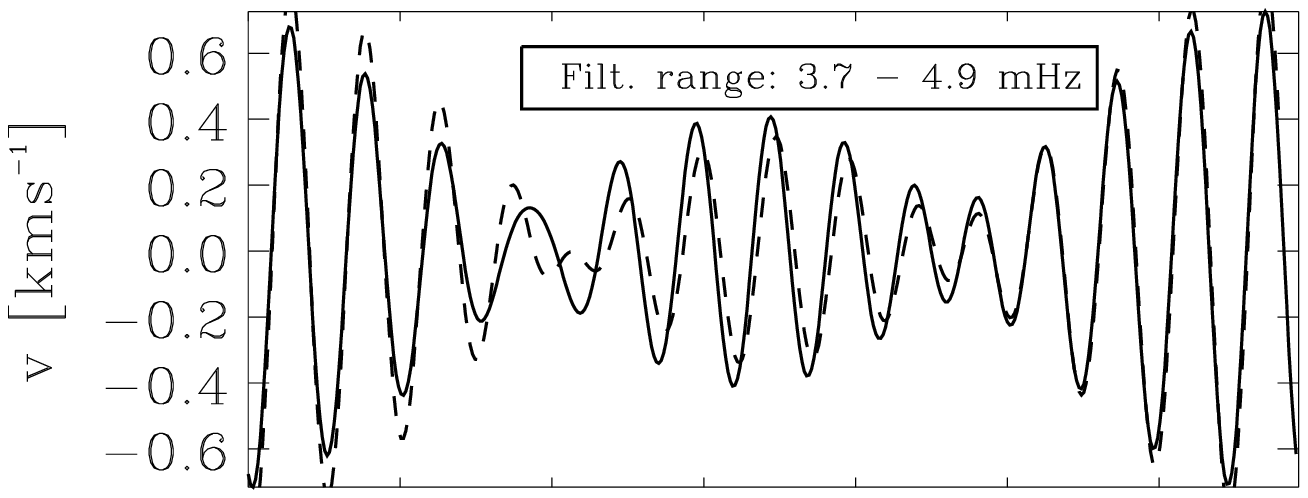} 
\includegraphics[width=9cm]{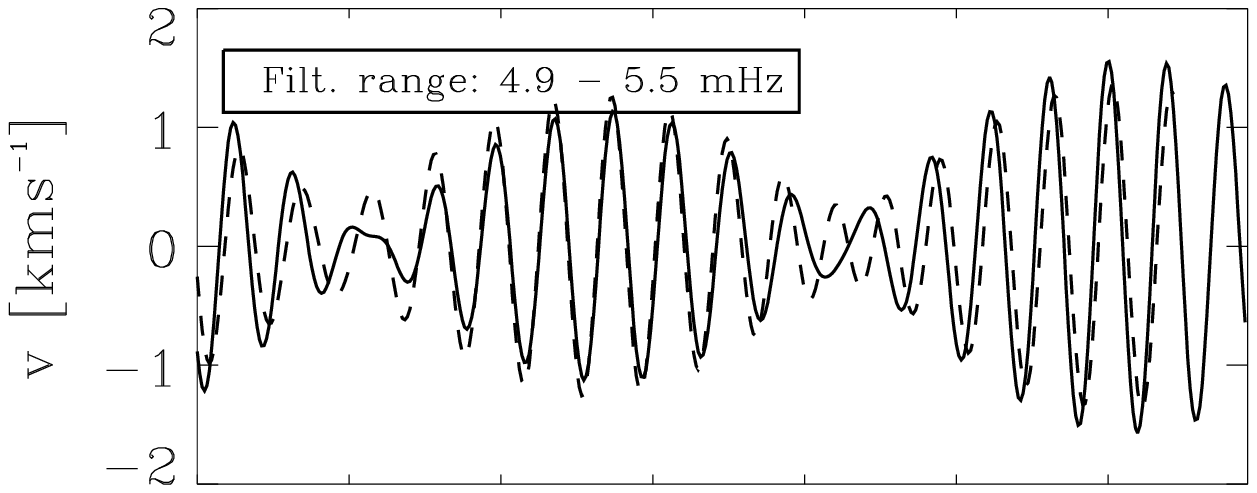} 
\includegraphics[width=9cm]{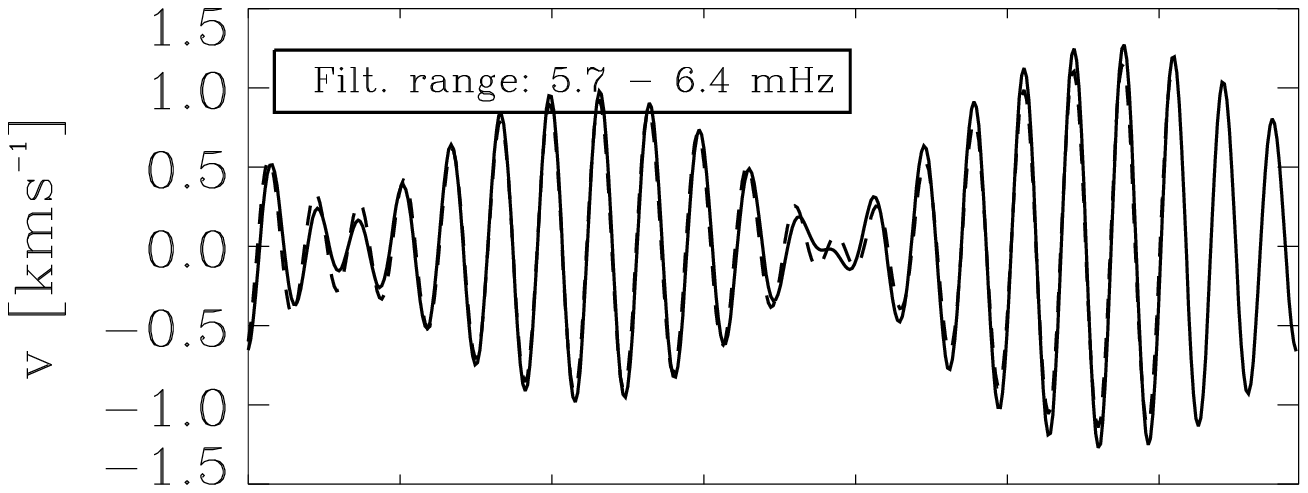} 
\includegraphics[width=9cm]{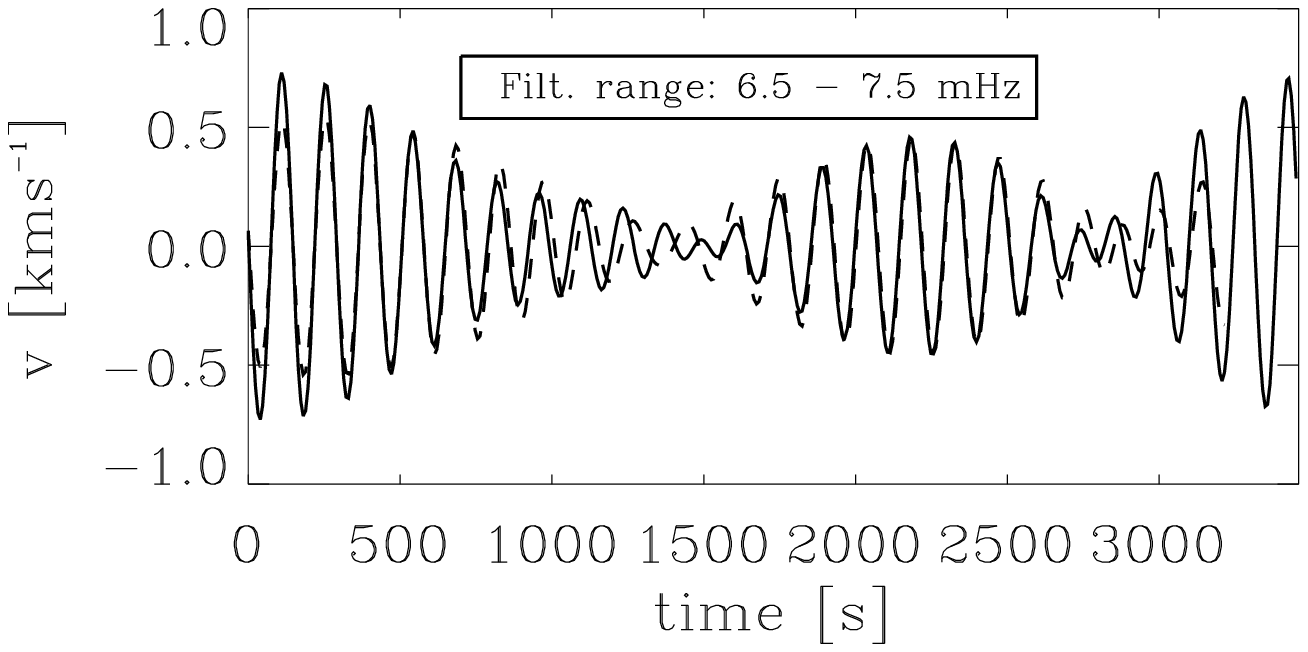} 
\caption{Time delays in the umbra of the small sunspot. 
The panels show the velocities, for one pixel position inside the umbra, filtered in 1 mHz frequency bands around 4, 5, 6 and 7 mHz. Overplotted to the 
chromospheric velocity (dashed lines) is the photospheric signal (solid lines), filtered
in the same frequency band, but amplified and delayed to make it match the latter one. Similar plots for other pixels along the slit result in delays and amplifications that are consistent 
with those shown here.}
\label{fig:time-delay}
\end{figure}

\begin{figure}[t!]
\center 
\includegraphics[width=9cm]{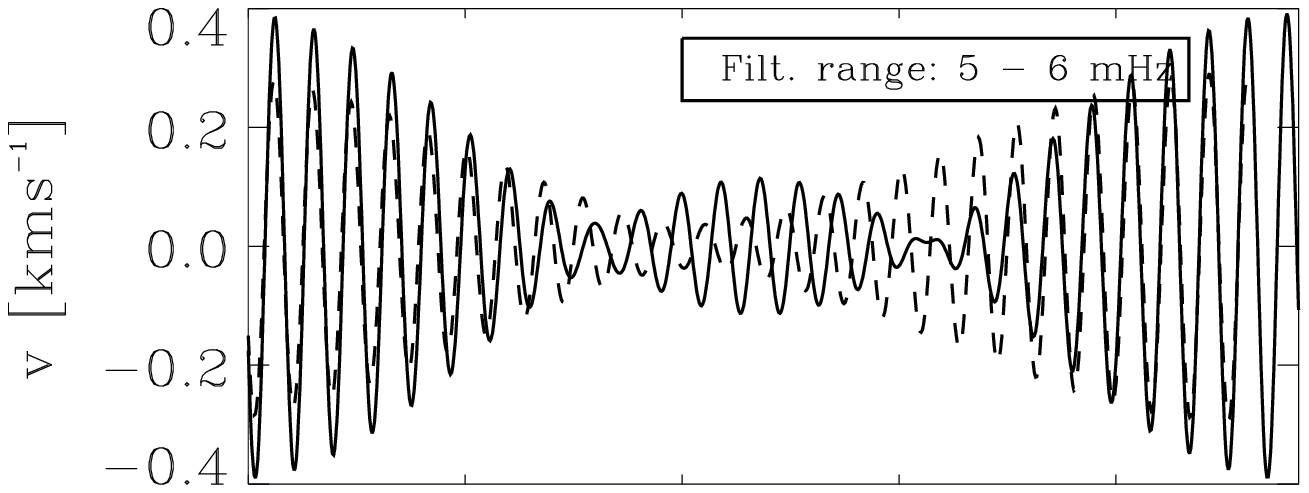} 
\includegraphics[width=9cm]{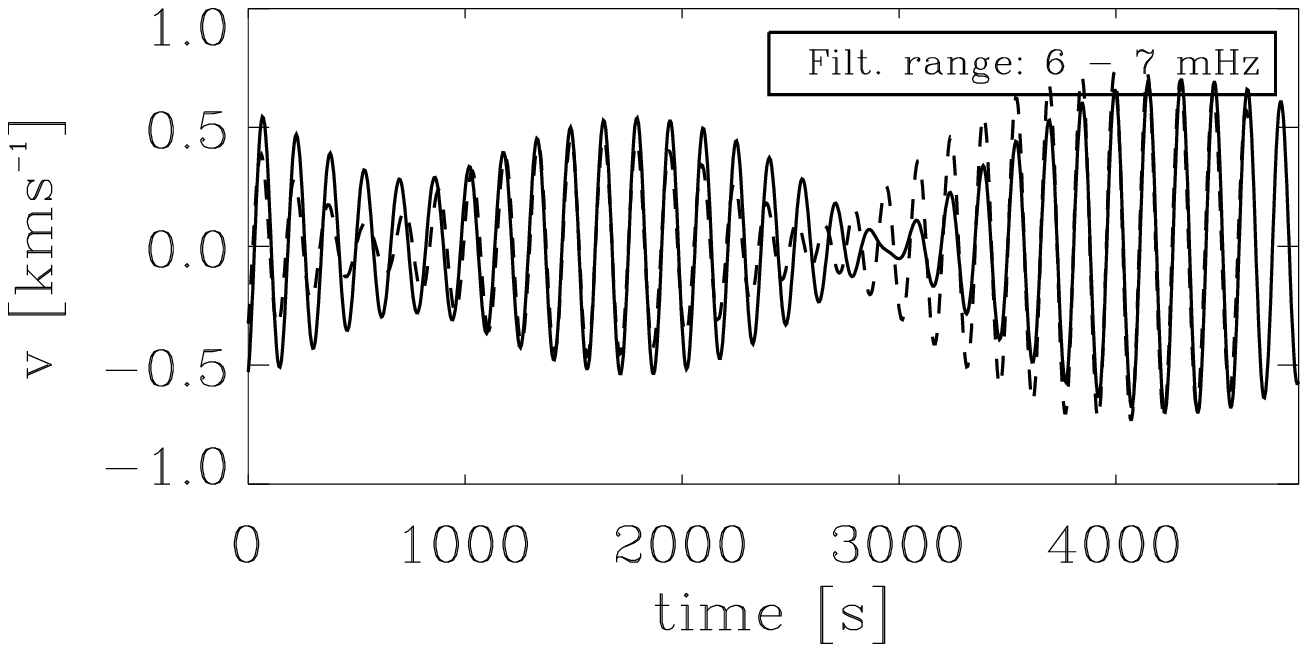}  
\caption{Measured time delays inside the pore. Analogous to 
Fig. \ref{fig:time-delay} except for the filtering bands. From top to bottom, the velocity
signals have been filtered in 1 mHz bands around 5.5 and 6.5 mHz, respectively. The photospheric signal was shifted in time and amplified in order to match the chromospheric one.}
\label{fig:pore-time-delay}
\end{figure}

\begin{figure}[t!]
\center 
\includegraphics[width=9cm]{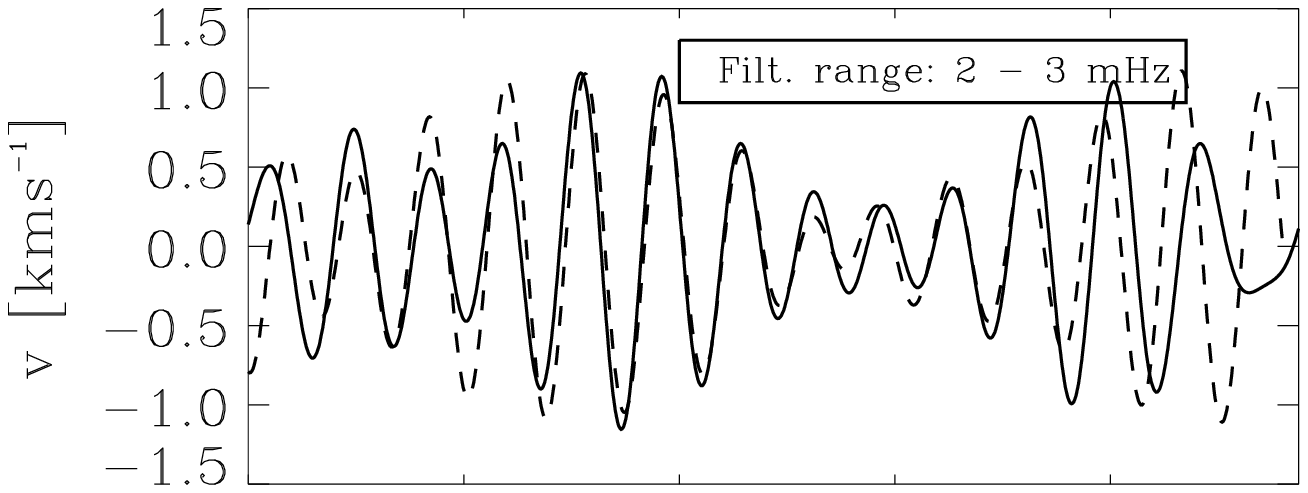} 
\includegraphics[width=9cm]{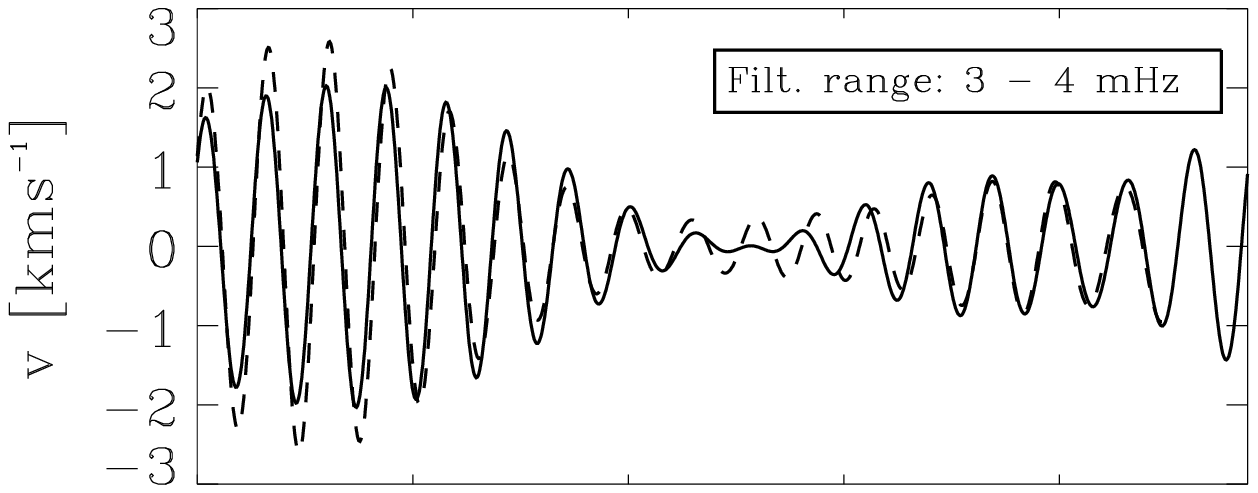} 
\includegraphics[width=9cm]{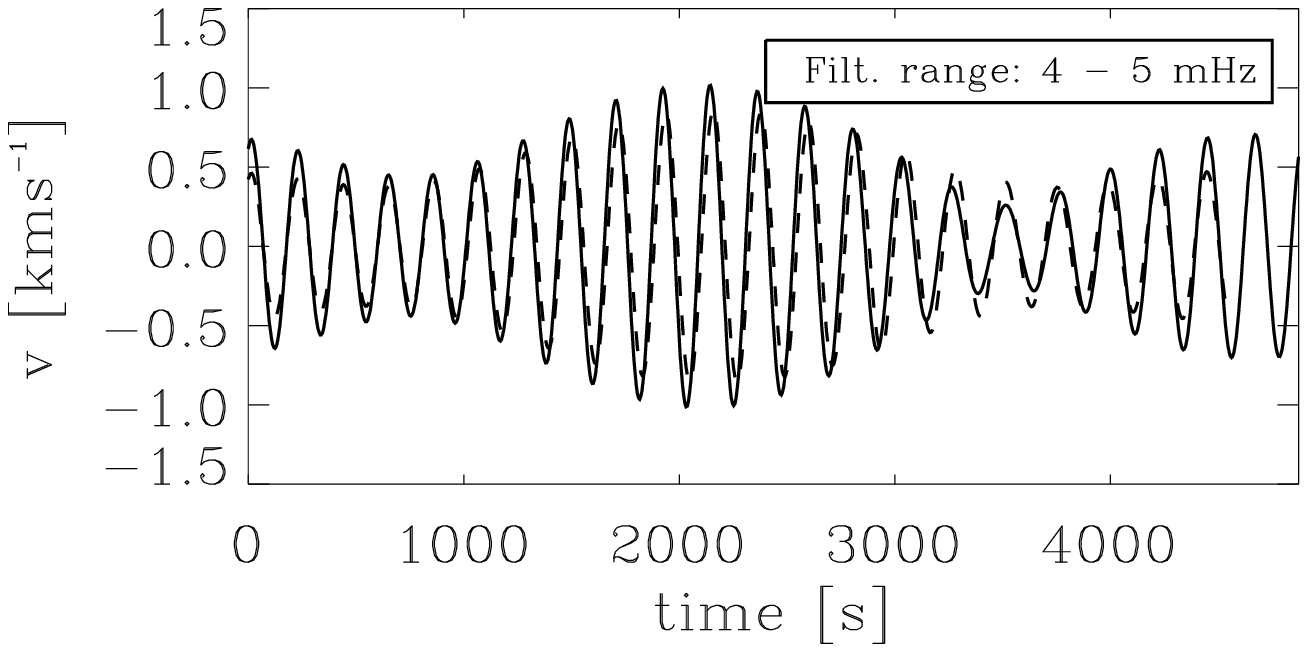} 
\caption{Measured time delays inside the facular region. Analogous to 
Fig. \ref{fig:time-delay} except for the filtering bands. From top to bottom, the velocity
signals have been filtered in 1 mHz bands around 2.5, 3.5 and 4.5 mHz, respectively. The photospheric signal was shifted in time and amplified in order to match the chromospheric one.}
\label{fig:facula-time-delay}
\end{figure}

\clearpage

The solid lines in  Fig.~\ref{fig:theor-time-delay} show the expected time delays 
(in the four magnetic structures), as a function of frequency, between the oscillations measured at the photosphere and the chromosphere as derived from the 
theoretical model. The time delay, $\Delta t$, depends on the difference in heights, $\Delta z$
and the group velocity of the wave packet, $v_g$:

\begin{equation}
\Delta t = \frac{\Delta z}{v_g}
\end{equation}

\noindent where 

\begin{equation}
v_g = \frac{d \omega}{d k_z}
\end{equation}

\noindent Using the best fitting values for $\Delta z$, $T$ and $\tau_R$, 
we computed the theoretical time delay from the wave propagation model described above.

\noindent If propagation is mainly linear within the 2 - 7 mHz band (as suggested by the good fits of the model to the data)
and it takes place along the magnetic field lines, then we should expect to see a 
correlation of the photospheric and chromospheric oscillation patterns above the 
cutoff-frequency.
It should be possible to determine the time delay from the observations by simply
comparing these modulation patterns at both heights. 
The time shift that yields the maximum correlation will correspond to the measured delay.

However, the theoretical time delays depicted in Fig. \ref{fig:theor-time-delay} show a very strong dependence on the 
frequency of the oscillating mode, so we have to take this fact into account
when doing the comparison. Following the approach taken in Paper I, we first filter 
both the photospheric and the chromospheric velocity maps in narrow 
frequency ranges (narrow enough that the expected time delay does not 
vary significantly within the bandwidth).
Then, we compare the photospheric and chromospheric filtered signals finding that
we have to apply a certain time shift between them in order to make their external
modulation schemes match. This shift corresponds to the time that a perturbation
(within the filtering frequency range) originated in at photospheric levels takes to reach
the chromosphere.

In the case of the sunspot-like structures, we filtered the velocity maps in 
several one-mHz bands close to 6 mHz (where the main contribution to the 
chromospheric power lies). 
We then compared each pair of filtered maps finding a time shift between them 
that depended on the frequency range in which the maps were filtered. 
The four panels of Fig. \ref{fig:time-delay} show the photospheric (solid) and 
chromospheric (dashed) velocities, 
for one position inside the umbra of the small sunspot, filtered in $\sim 1$ mHz
frequency bands around 4, 5, 6 and 7 mHz. In each case, the photospheric signal has been amplified 
and delayed to make it match the chromospheric one.

\noindent This procedure was repeated for several pixels in the umbra.
There are coherent patches of a few arcseconds along the slit in 
which the velocity signals are very similar, so we chose 3 or 4 
positions far apart, and measured the delay for each of them finding consistent 
time shifts and amplification values at the different spatial locations. The 
uncertainty for the time delay is of the order of 1 - 2 time steps ($\sim$ 10 - 15s).
The measured delays extracted from this method are represented by the asterisks
over plotted to the top left panel of Fig. \ref{fig:theor-time-delay}. Analogous analyses
were carried out for the pore (shown in Fig. \ref{fig:pore-time-delay}) and for the big 
sunspot (in Paper I).

\noindent The pore is only 4 arcsecs wide and oscillations are quite 
coherent throughout the whole structure. Also, there is not a strong 
modulation of the velocity oscillation pattern, so, in certain 
frequency bands, it is not clear what time shift gives the larger 
correlation between photosphere and chromosphere. This is why we only show 
the results for two frequency bands in Fig. \ref{fig:pore-time-delay}.

\noindent In the case of the facular region we filtered the velocity maps 
in three ranges around 3 mHz (where both photospheric and chromospheric power spectra have their main contributions). The remaining analysis is parallel to the former 
case. The three panels of Fig. \ref{fig:facula-time-delay} show the chromospheric 
({\em solid}) and photospheric ({\em dashed}) velocity signals 
filtered around 2.5, 3.5 and 4.5 mHz, respectively. Again, the photospheric signal has
been amplified and shifted in time to make it match the chromospheric one.

In all the cases, the amplification factors turn out to be consistent with
the values of the amplification spectrum in Fig. \ref{fig:phase},
and the measured time delays (represented by asterisks superposed to the theoretical 
time delays in Fig. \ref{fig:theor-time-delay}) obtained from the shifts are 
consistent with the expected values obtained from the model. 
Even though the theoretical curves predict a very strong variation of the time 
delay within the 1 mHz filtering bands, the measured delays agree surprisingly well 
with what is expected by the model.

\section{Discussion and conclusions}

In this paper we have investigated the wave propagation in the atmospheres of four solar 
magnetic structures with decreasing
flux (two different-sized sunspots, a pore and a facular region). 
Simultaneous and co-spatial measurements of the LOS velocity at the photosphere 
and the chromosphere of these structures allow us to infer information about 
the properties of wave propagation from one atmospheric layer to another. A simple model of 
linear vertical wave propagation in a magnetized stratified medium with radiative 
losses is enough to explain the observed phase difference and amplification spectra 
and the measured time delays.

The inversion of the full Stokes vector in the four structures reveals vertical (i.e. radial)
magnetic fields in all cases. The comparison of the photospheric and chromospheric velocity
maps, filtered in narrow frequency ranges, shows a pixel to pixel correlation along the slit 
of the external modulation of the 
wave pattern (after accounting for a global time shift and a global amplification of the 
photospheric signal). These two facts are enough to justify the election of a model of
linear wave propagation along vertical magnetic field lines.

In the case of sunspot-like structures the atmospheric cut-off lies around $\sim$4 mHz,
so the modes with frequencies below this one will not be able to reach the high 
chromosphere. Many authors have argued before about the possibility 
of the non-linear interaction among 5-min modes being the source of the
chromospheric 3-min oscillations \citep[see, e.g.][]{fleck}. But if this
were the case, the photospheric and chromospheric filtered velocity maps would show no
resemblance with each other. As a matter of fact, a clear correlation exists, 
indicating that most of the 6 mHz power observed at chromospheric heights 
in sunspot-like structures comes directly from the same frequency range in the
photosphere via upward linear wave propagation. This ratifies and extends the
conclusions of Paper I (which focuses on the umbrae of big sunspots) to smaller sunspots and
pores. As the size of the structure decreases, so does the typical radiative cooling time,
while the temperature, on the other hand, grows. As argued before, both these 
behaviors are in agreement with what is expected from a qualitative point of view.

In the case of facular regions, the cut-off frequency stands around 2 mHz.
This is accounted for in the model by introducing a shorter cooling time and a higher
temperature. This allows the 5-min power to
propagate through the atmosphere and reach the high chromosphere. Again, a clear correlation between photospheric and chromospheric
filtered velocity maps can be found, indicating that the propagation is mainly linear and vertical
within the 2--5 mHz range.

\noindent It is interesting to point out that, in this particular case, there is no need to 
invoke a large inclination of magnetic flux tubes to explain the p-mode leakage into the chromosphere \cite[][]{depontieu2004}. 
Furthermore, the comparison of the photospheric and chromospheric velocity maps shows a good 
co-spatial correlation (after applying a convenient time shift) indicating that
the propagation is essentially vertical. If we assume that the wave propagation takes place
along the field lines and we take into account a height difference of 1000 - 1500 km,
a magnetic field inclination of $40 - 45^{\circ}$ would result in a spatial displacement of
3 - 5 pixels between the photospheric and the chromospheric oscillations (rather than 
happening on the same vertical). In the best case scenario, this displacement would be parallel to
the slit; however, in the worst case, it would happen along the direction perpendicular to
the slit, leading to a correlation between the photosphere and the chromosphere that is
marginally compatible with the spatial resolution of
our data. On the other hand, the measured Stokes profiles set an upper limit of 
$20^{\circ}$ on the inclination of the photospheric facular magnetic fields. Both these arguments point
towards a magnetic field structure that is incompatible with the inclinations needed to lower 
the cut-off frequency enough to allow p-mode leakage.

\noindent The numerical simulations of Khomenko et al. (2008) have confirmed this 
conclusion since they show how, in a more realistic atmosphere, it is possible to explain the propagation of 5-minute modes into 
the chromosphere through vertical thin flux tubes, using radiative losses as the
main ingredient to lower the cut-off frequency.

As the photospheric perturbations propagate upwards, their amplitude
increases due to the rapid decrease in density and they eventually develop
asymmetries that steepen more or less depending on the magnetic structure
(the lower the magnetic flux, the smaller the amplification and the steepness of the developing shock).
In all the cases, the time delay between photospheric and chromospheric
oscillations is around several minutes in the frequency range near where the
chromospheric power peaks.

\acknowledgments
This research was partially funded by the Spanish Ministerio de Educaci\'on 
y Ciencia through project AYA2007-63881.
The National Center for Atmospheric Research (NCAR) is sponsored by the 
National Science Foundation

\end{document}